\documentclass[twocolumn]{aastex631}

\usepackage{newtxtext,newtxmath}
\usepackage{ae,aecompl}

\usepackage{graphicx}	
\usepackage{amsmath}	

\usepackage[dvipsnames]{xcolor}
\usepackage{multirow}
\usepackage{amssymb}
\usepackage{natbib}

\defcitealias{HaardtMadau2012}{HM12}
\defcitealias{Puchwein_2019}{P19}
\defcitealias{Faucher_Gigu_re_2020}{FG20}
\defcitealias{FaucherGiguere2009}{FG09}
\defcitealias{HaardtMadau2001}{HM01}

\newcommand\msun{{\,M_\odot}}

\newcommand{\cmci}{~\mbox{cm}^{-3}}

\newcommand{\K}{~\mbox{K}}

\newcommand{\traphic}{{\sc traphic}}

\newcommand{\gadgetthree}{{\sc p-gadget3}}

\newcommand{\Msun}{M_{\odot}}

\newcommand{\kmsMpc}{~\mbox{km~s}^{-1}~\mbox{Mpc}^{-1}}

\newcommand{\music}{{\sc music}}
\newcommand{\cloudy}{{\sc cloudy}}
\newcommand{\rockstar}{{\sc rockstar}}

\newcommand{\st}{{\sc starburst99}}

\newcommand{\universemachine}{{\sc UniverseMachine}}

\newcommand{\nspace}{{ }}

\newcommand{\BG}{\textsc{BG}}
\newcommand{\Rad}{\textsc{Rad}}

\begin{document}

\title{Patchy Reionization Delays Quenching in Isolated Ultra-faint Dwarf Galaxies}

\author[0009-0004-0053-064X]{Juyoung Kim}
\affiliation{School of Space Research, Kyung Hee University, 1732 Deogyeong-daero, Yongin-si, Gyeonggi-do 17104, Republic of Korea}

\author[0000-0001-6529-9777]{Myoungwon Jeon}
\affiliation{School of Space Research, Kyung Hee University, 1732 Deogyeong-daero, Yongin-si, Gyeonggi-do 17104, Republic of Korea}
\affiliation{Department of Astronomy \& Space Science, Kyung Hee University, 1732 Deogyeong-daero, Yongin-si, Gyeonggi-do 17104, Republic of Korea}

\author[0000-0001-7863-2591]{Yumi Choi}
\affiliation{NSF National Optical-Infrared Astronomy Research Laboratory, 950 North Cherry Avenue, Tucson, AZ 85719, USA}

\author[0000-0001-8368-0221]{Sangmo Tony Sohn}
\affiliation{Space Telescope Science Institute, 3700 San Martin Drive, Baltimore, MD 21218, USA}
\affiliation{Department of Astronomy \& Space Science, Kyung Hee University, 1732 Deogyeong-daero, Yongin-si, Gyeonggi-do 17104, Republic of Korea}

\correspondingauthor{Myoungwon Jeon}
\email{myjeon@khu.ac.kr}

\submitjournal{ApJ}
\received{May 25, 2026}
\revised{August 23, 2026}
\accepted{September 14, 2026}

\begin{abstract}
The spatially inhomogeneous nature of cosmic reionization is directly encoded in the quenching timescales of ultra-faint dwarf galaxies (UFDs). While most cosmological simulations employ a spatially uniform UV background (UVB) for computational efficiency, such homogeneous treatment may oversimplify the physical impact of reionization on UFDs, whose star formation histories (SFHs) are highly sensitive to local environments. Under these idealized, instantaneous UVB frameworks, star formation in UFDs is typically truncated abruptly at the onset of reionization, creating a tension with the extended and diverse SFHs observed in Local Group UFDs. To address this discrepancy, we perform high-resolution cosmological zoom-in simulations to contrast two distinct reionization models: a conventional uniform UVB and an on-the-fly radiative transfer (RT)-based patchy reionization model. We find that RT-based reionization delays star-formation quenching by $\sim 440\,\rm{Myr}$ on average, a result consistent with recent observational constraints. However, this extended star formation does not necessarily leave a discernible imprint on the stellar metallicity distribution, indicating that the delay—while physically significant for gas dynamics—is still insufficient to drive substantial chemical enrichment. Our results emphasize the necessity of more realistic, patchy reionization modeling to explain the observed diversity of the Local Group population.
\end{abstract}


\section{Introduction}
Cosmic reionization represents a fundamental phase transition in the early Universe, during which the intergalactic medium (IGM) was transformed from a cold, neutral state to a warm, ionized phase. This transition was initiated by the escape of ionizing photons from the first generations of stars and galaxies, and was subsequently driven by growing populations of low-mass galaxies (e.g., \citealp{BarkanaLoeb2001}; \citealp{Fan2006}; \citealp{LoebFurlanetto2013}; \citealp{Robertson2015}), with possible additional contributions from active galactic nuclei (AGNs; e.g., \citealp{MadauHaardt2015, Madau2024}). On galactic scales, this global phase transition played a critical role in regulating star formation, thus shaping the mass assembly history and subsequent evolution of early stellar populations.

In many state-of-the-art cosmological hydrodynamical simulations, such as EAGLE (\citealp{Crain2015EAGLE}), SIMBA (\citealp{Dave2019SIMBA}), IllustrisTNG (\citealp{Nelson2019TNG}), and FIREbox (\citealp{Feldmann2023FIREbox}), the reionization effect is conventionally implemented via a spatially homogeneous meta-galactic UV/X-ray background (UVB), employing tabulated, redshift-dependent radiation fields (e.g. \citealp{HaardtMadau2001}, hereafter \citetalias{HaardtMadau2001}; \citealp{HaardtMadau2012}, hereafter \citetalias{HaardtMadau2012}; \citealp{Puchwein_2019}, hereafter \citetalias{Puchwein_2019}; \citealp{FaucherGiguere2009}, hereafter \citetalias{FaucherGiguere2009}; \citealp{Faucher_Gigu_re_2020}, hereafter \citetalias{Faucher_Gigu_re_2020}).

However, by construction, these uniform models neglect the discrete distribution of radiation sources and the resulting ``patchy" topology of reionization (\citealp{Becker2015}), which is particularly critical for low-mass systems such as ultra-faint dwarf galaxies (UFDs, $M_\star < 10^{5}\Msun$ and $L < 10^{5} L_\odot$) (see the review by \citealp{Simon2019}). Inherently, reionization is patchy, beginning earliest in high-density regions near ionizing sources and completing latest in the underdense voids farthest from them. Consequently, such localized environmental variations determine the thermal evolution of the gas and govern the duty cycle of star formation, particularly in low-mass halos  (e.g., \citealp{Iliev_2014}; \citealp{Aloisio2015}; \citealp{Qin2021}; \citealp{GnedinMadau2022_LRCA}). 


Despite this, the majority of UFD simulations have employed spatially uniform UVBs (e.g., \citealp{Wheeler2015}; \citealp{Jeon_2017}; \citealp{Ma2017}; \citealp{Revaz2018}; \citealp{Sanati2023}; \citealp{Hopkins2023})—meaning that the radiation field varies only with redshift but is applied identically across the simulation volume without spatial variations or photon-propagation effects. This reliance stems from the numerical challenge of achieving sufficient resolution to resolve UFD-scale halos while simultaneously performing full-radiative transfer (RT) within a large-scale cosmological volume. Interestingly, under the widely adopted ``flash-like" UVB implementation, star formation in UFDs is likely to be immediately quenched once the UVB is switched on (e.g., \citealp{Wheeler2015}; \citealp{Jeon_2017}; \citealp{Ma2017}; \citealp{Kim2023}). 

On the other hand, observational evidence reveals that Local Group (LG) UFDs exhibit a broad spectrum of star formation histories (SFHs), ranging from sharply truncated to significantly prolonged star formation (e.g., \citealp{Weisz_2014, Brown_2014, Weisz_2017, Weisz2019, Sacchi_2021, Savino2023, Meredith2025}). This observed diversity stands in contrast to the synchronous quenching typically predicted by theoretical models that adopt a flash-like UVB implementation.
Such diversity in the observed SFHs could be driven by a complex interplay of internal physical properties, including halo mass, mass assembly history, and the physical state of their gas reservoirs. Alternatively, the specific timing at which a system is exposed to the UVB can serve as another critical driver. Some observational studies support that environmental variations and the resulting localized reionization effects potentially contribute to this evolutionary diversity (see, e.g., \citealp{Sacchi_2021, Meredith2025}).

For example, \citet{Sacchi_2021} derived HST-based SFHs for seven UFDs and found that candidate satellites of the Large Magellanic Cloud (LMC) were quenched, on average, $\sim600\,{\rm Myr}$ later than other MW UFDs. This discrepancy may be attributed to the timing of reionization; LMC satellites likely experienced reionization later than MW satellites, which were situated within the host environment earlier. Recently, using a significantly larger sample, \citet{Meredith2025} derived SFHs for 36 UFDs within the MW halo, demonstrating that while their overall quenching timescales are compatible with reionization-driven suppression, LMC satellites were quenched up to $\sim800\,{\rm Myr}$ later than long-term MW satellites. Such observational constraints emphasize the necessity for UFD-scale simulations that explicitly incorporate patchy reionization to capture the localized physical processes governing UFD evolution.

To account for the impact of patchy reionization on UFDs, \cite{Kim2023} addressed the limitations of the standard homogeneous meta-galactic UVB (\citetalias{HaardtMadau2012}) by constructing an environment-dependent, yet spatially uniform, UVB model. In their approach, the ionizing radiation field specific to a given simulated UFD was obtained by summing the spectral energy distributions (SEDs) of surrounding ionizing sources. The resulting UVB was then tabulated as a function of redshift and imposed on the hydrodynamic simulations. With this prescription, the quenching times of UFD analogs were delayed by $\sim460\,{\rm Myr}$ compared to runs adopting the \citetalias{HaardtMadau2012} model, indicating that an environment-dependent reionization history can substantially extend star formation in low-mass systems. However, because reionization was still implemented via a ``flash-like"  approximation, these simulations remained limited in their ability to explicitly capture the time-dependent propagation of ionization fronts.

This missing time-dependent propagation has been addressed primarily in large-volume RT simulations, such as CROC (\citealp{Gnedin2014}), CoDa (\citealp{Ocvirk2016}), and THESAN (\citealp{Kannan2022}), which explicitly follow the transport of ionizing photons across cosmological volumes. These studies have shown that reionization strongly affects low-mass halos, particularly below $10^{9} \Msun$, by suppressing gas accretion and star formation. This picture is consistent with observational constraints (\citealp{Weisz_2014, Brown_2014, Weisz_2017}) and high-resolution zoom-in simulations adopting uniform UVBs (\citealp{Wheeler2015, Sawala2016, Jeon_2017, Simpson_2018, Wheeler2019, GarrisonKimmel2019, Rey2020, Applebaum2021}).  However, because resolving UFD-scale systems requires computationally demanding mass resolution, large-volume simulations lack the small-scale details, leaving the impact of full RT-based patchy reionization on UFD evolution largely unexplored within high-resolution cosmological zoom-in simulations.

In this work, we implement RT-based patchy reionization in high-resolution zoom-in simulations of UFDs. Our main goal is to examine how a more realistic modeling of patchy reionization affects the formation and evolution of UFDs. To this end, we conduct a comparative analysis using two distinct prescriptions derived from the same ionizing sources. The first is a spatially uniform UVB model, and the second is an RT-based model that explicitly tracks the time-dependent propagation of ionizing radiation.
We follow the evolution of these systems down to $z=5$, the epoch by which hydrogen reionization is observed to be complete (e.g., \citealp{Becker2015, Mcgreer2015, Bosman2022}). By evaluating the resulting gas and stellar properties, we assess how reionization modeling governs the physical properties of the simulated UFDs. In particular, we investigate whether the RT-based approach facilitates more prolonged SFHs compared to uniform UVB models. Finally, we identify which observable quantities preserve the structural signatures of such extended SFHs.

The paper is organized as follows. In Section \ref{Sec:Method}, we describe the simulation methodology, including the treatment of star formation, stellar feedback, and the two reionization schemes adopted in this work. In Section \ref{Sec:Result}, we present the results, focusing on the gas and stellar properties of the simulated UFDs. In Section \ref{Sec:Discussion}, we discuss the implications of our results for previous UFD simulations and highlight several caveats. Finally, in Section \ref{Sec:Summary}, we summarize the main findings of this study.

\section{Numerical methodology}
\label{Sec:Method}
\subsection{Simulation setup}
We perform cosmological hydrodynamic simulations using a modified version of the N-body smoothed particle hydrodynamics (SPH) code \gadgetthree \nspace (\citealp{2001NewA....6...79S}; \citealp{Springel_2005}). For the cosmological parameters, we adopt a matter density parameter of $\Omega_{\rm m} = 1 - \Omega_{\rm \Lambda} = 0.265$, a baryon density of $\Omega_{\rm b} = 0.0448$, a Hubble constant of $H_{\rm 0} = 71 \kmsMpc$, a spectral index of $n_{\rm s} = 0.963$, and a normalization of $\sigma_{\rm 8} = 0.8$ (\citealp{Komatsu2011}; \citealp{Planck2016}). The initial conditions of the simulations are generated from the cosmological initial condition codes \music \nspace (\citealp{Hahn_2011}).
\label{Simulation setup}

\par
To identify target halos at $z = 0$, we conduct a dark matter (DM)-only simulation with $256^{3}$ particles in a $L_{\rm {box}} = 6.25h^{-1}$ $\rm{cMpc}$. The target halos are then identified using the halo finder codes \rockstar \nspace (\citealp{Behroozi_2013}). After the halo finding procedure, we select seven halos with a mass of $10^{8} \lesssim M_{\mathrm{vir}}/M_{\odot} \lesssim 10^{9}$ at $z = 0$. These seven halos are isolated, since we only choose halos without another halo with masses $M_{\mathrm{vir}}>M_{\mathrm{vir,\ target}}$ within $10 R_{\rm vir,\ target}$ (\citealp{Agertz2019}), which helps reduce computational cost by avoiding additional massive structures in the vicinity.

\par
Next, we apply the zoom-in technique with four successive refinements to the regions enclosing the area $2.5$ times the virial radius of the target UFD halo at $z = 0$. The final resolution of the most refined region is $4096^{3}$, with DM and gas particle masses of $m_{\rm DM} \approx 300 {\rm \msun}$ and ${m_{\rm gas} \approx 63 {\rm \msun}}$, respectively. We adopt the softening length value of $\epsilon_{\rm soft} \sim 30$ pc, constant for DM and stellar particles at all simulation times. However, for gas particles, we utilize a flexible softening length proportional to the length of the SPH kernel with a minimum value of $\epsilon_{\rm gas,\,min} = 2.8$ ${\rm pc}$.

\par
In every simulation timestep, we solve the non-equilibrium rate equations to update the primordial chemistry of atomic and molecular species, H, H$^+$, H$^{-}$, H$_2$, H$^{+}_{2}$, He, He$^+$, He$^{++}$, e$^{-}$, D, and D$^+$. Furthermore, not only primordial cooling, our simulations include metal cooling processes involving seven species, C, N, O, Mg, Si, Ne, and Fe. We determine the cooling rates for these elements based on photoionization codes \cloudy \nspace (\citealp{1998PASP..110..761F}).

\subsection{Star formation}
In our simulations, stars are formed when the hydrogen number density of a gas particle exceeds the density threshold of $n_{\rm H,\,thr} = 10^{2}\,{\rm cm}^{-3}$ and has a temperature lower than the threshold temperature $T_{\rm {thr}}=10^{4}\,{\rm K}$. The conversion of a gas particle to a star particle is stochastic, following the conversion rate from the Schmidt law (\citealp{Schmidt1959}). In particular, the star formation rate is governed by the equation $\dot{\rho}^{*} = \rho / \tau_{\star}$, where $\rho$ represents the gas density, and $\tau_{\star}$ is the timescale of star formation. Here, $\tau_{\star}$ is given by $\tau_{\star} = \tau_{\rm ff} / \epsilon_{\rm ff}$, where $\tau_{\rm ff}$ is the free-fall time at the gas density $\rho$ and $\epsilon_{\rm ff}$ is the star formation efficiency. The free-fall time $\tau_{\rm ff}$ is defined as $\tau_{\rm ff} = [3\pi/(32G\rho)]^{1/2}$. We set the star formation efficiency $\epsilon_{\rm ff} = 0.01$ for both Population III (Pop III) and Population II (Pop II) stars in this study, which is a representative value observed in the local Universe (e.g., \citealp{Leroy2008}). The resultant star formation timescale $\tau_{\star}$ is as follows,
\begin{equation}
\tau_{\star} = \frac{\tau_{\rm ff}(n)}{\epsilon_{\rm ff}} \sim 400 \, \rm Myr \, \it \, \Big(\frac{n}{\rm 10^{2}  \cmci}\Big)^{\rm -1/2} .
\end{equation}
\label{starformation}

\par
We categorize the stars into Pop III and Pop II based on a critical metallicity of $Z_{\rm crit} = 10^{-5.5}\,Z_{\odot}$. Gas with metallicity below this threshold forms Pop III stars, whereas gas exceeding it forms Pop II stars (e.g., \citealp{Omukai2000}; \citealp{Schneider2010}; \citealp{Safranek-Shrader2016}).

\subsubsection{Pop III star}
Pop III stars form in pristine environments where molecular hydrogen ($\rm{H_{2}}$) cooling, being less efficient than metal-line cooling, leads to a systematically more massive population (e.g., \citealp{Bromm2013, KlessenGlover2023}). Despite uncertainties in their mass range (e.g., \citealp{Bromm2013}; \citealp{Hirano2014}; \citealp{Hirano2015}; \citealp{KlessenGlover2023}), our high mass resolution enables us to model Pop III stars as individual star particles. We sample individual stars from the top-heavy initial mass function (IMF) over the mass range $[1, 260]\, \Msun$, following the equation below,
\begin{equation}
\phi=\frac{dN_{\rm {Pop\,III}}}{d\,{\rm log}\,m} \propto m^{-1.3}\exp\left[-(\frac{m_{\rm char}}{m})^{1.6}\right],
\end{equation}
where the characteristic mass $m_{\rm char}$, is $40\Msun$ (e.g. \citealp{Wise2012}). When a gas particle of mass $63\Msun$, corresponding to our gas particle mass resolution, is converted into a Pop III star, we allow a small mismatch between the gas mass removed and the stellar mass assigned, rather than enforcing exact mass conservation. This has a negligible impact because Pop III formation occurs rarely compared to Pop II, and enrichment from just one or two Pop III SNe is sufficient to initiate the Pop II star formation that dominates the system thereafter.

\subsubsection{Pop II star}
\label{Sec:starformationPopII}
For Pop II stars, we sample stellar masses from a Salpeter IMF (\citealp{Salpeter1955}) over $0.1-100\Msun$, following the equation.

\begin{equation}
\phi=\frac{dN_{\rm Pop\,II}}{d\,{\rm log}\,m} \approx m^{-\alpha},
\end{equation}
where the slope $\alpha=1.35$.
\par
If the sampled star mass is between $0.1\Msun$ and $8\Msun$, we continue the sampling process until their cumulative sum reaches $63\Msun$. Once it happens, we stop the sampling and consider the resulting collection as a single stellar population (SSP), hereafter Pop II-SSP. Otherwise, if the sampled mass is between $8\Msun$ and $100\Msun$, we regard it as an individual star particle, hereafter Pop II-indiv. These massive stars undergo SNe or collapse directly into black holes (BHs), according to their sampled mass (for details, see \ref{PopIIfeedback}). Since massive individual stars collapse into SNe or BHs, we only include SSP particles for stellar mass and stellar metallicity analysis. For details of our Pop II star formation implementation, we refer the readers to \cite{Go2025} and \cite{JeonKo2026}.
\label{PopIIformation} 

\subsection{Stellar feedback}
\label{STELLARFEEDBACK}
\subsubsection{Photoionization feedback}
In this study, to capture the photon propagation mechanism, we use the RT module for the SPH code \traphic\nspace (\citealp{Pawlik_2008}). Briefly, \traphic\nspace solves the RT equation by tracing photon packets directly on the adaptive unstructured distribution of SPH particles in an explicitly photon-conserving manner. For details, we refer the reader to \cite{Pawlik_2008}. We consider photoionization feedback only from Pop III stars, which are expected to dominate the ionizing budget at the earliest stages of galaxy formation in our simulations, and neglect Pop II RT for computational efficiency; the Pop II contribution to reionization itself enters through the external sources (Section \ref{CR}). The amount of photons emitted from those Pop III stars is determined according to their initial mass. Specifically, we adopt polynomial fits from \cite{Schaerer2002}, ${\rm{log}_{10}} \dot{N}_{\rm ion} = 43.61 + 4.90x - 0.83x^{2}$, where $x = {\rm log_{10}} (m/\Msun)$ is the initial mass (For details, see \citealp{Jeon_2015}). We set the main sequence lifetime of Pop III stars as $t_{\rm Pop\, III} = \rm{3\,Myr}$, regardless of their initial mass.

\subsubsection{SN Feedback from Pop III stars} 
After their main-sequence lifetimes, massive Pop III stars explode as core-collapse SNe (CCSNe) or pair-instability SNe (PISNe) depending on their initial masses: $10\Msun \lesssim m \lesssim 40\Msun$ for CCSNe and $140\Msun \lesssim m \lesssim 260\Msun$ for PISNe. These explosions inject not only energy but also synthesized metals into the surrounding gas (e.g., \citealp{Heger2002}; \citealp{Heger2010}).

\par
In this work, SN feedback is implemented by depositing SN energy as thermal energy into the surrounding medium. To mitigate the well-known over-cooling problem associated with purely thermal injection, we adopt the scheme proposed by \citet{DallaVecchia2012}, which limits the number of neighboring gas particles receiving SN energy. For CCSNe, we inject $10^{51}\,\rm{erg}$ of energy into the single nearest gas particle, while for PISNe, $10^{52}\,\rm{erg}$ is distributed among the ten nearest gas particles. For SN enrichment, we adopt the nucleosynthetic yields and remnant masses of \cite{Heger2010} for CCSNe and \cite{Heger2002} for PISNe. For details of the implementation of chemical feedback and metal diffusion, we refer the readers to \cite{Jeon_2017}.

\subsubsection{SN Feedback from Pop II stars}
\label{PopIIfeedback}
As described in Section \ref{PopIIformation}, we distinguish two types of Pop II stars based on sampled masses: Pop II-SSP and Pop II-indiv. Once Pop II-indiv stars with masses between $8-40\Msun$ form, they explode as CCSNe with no delay time. For Pop II stars, we deposit $10^{51}\, \rm{erg}$ of energy into the single nearest gas particle and adopt the metal yields from \cite{Portinari1998}. Pop II-indiv stars with initial masses of $40-100\Msun$ are assumed to undergo direct collapse to BHs, injecting no radiative or SN feedback. From here on, to refer only to stars that provide feedback, we use ``Pop II-indiv'' exclusively for Pop II stars in the $8-40 \, \Msun$ mass range that explode as CCSNe. For Pop II-SSP stars, we include chemical enrichment from AGB mass loss and Type Ia SNe according to the mass distribution of the constituent stars in each SSP. For details of Pop II chemical feedback and metal diffusion, we refer the reader to \cite{JeonKo2026} and \cite{Jeon_2017}, respectively.

\subsection{Cosmic reionization}
\label{CR}
\begin{figure*}
     \centering
     \includegraphics[width = 175mm]{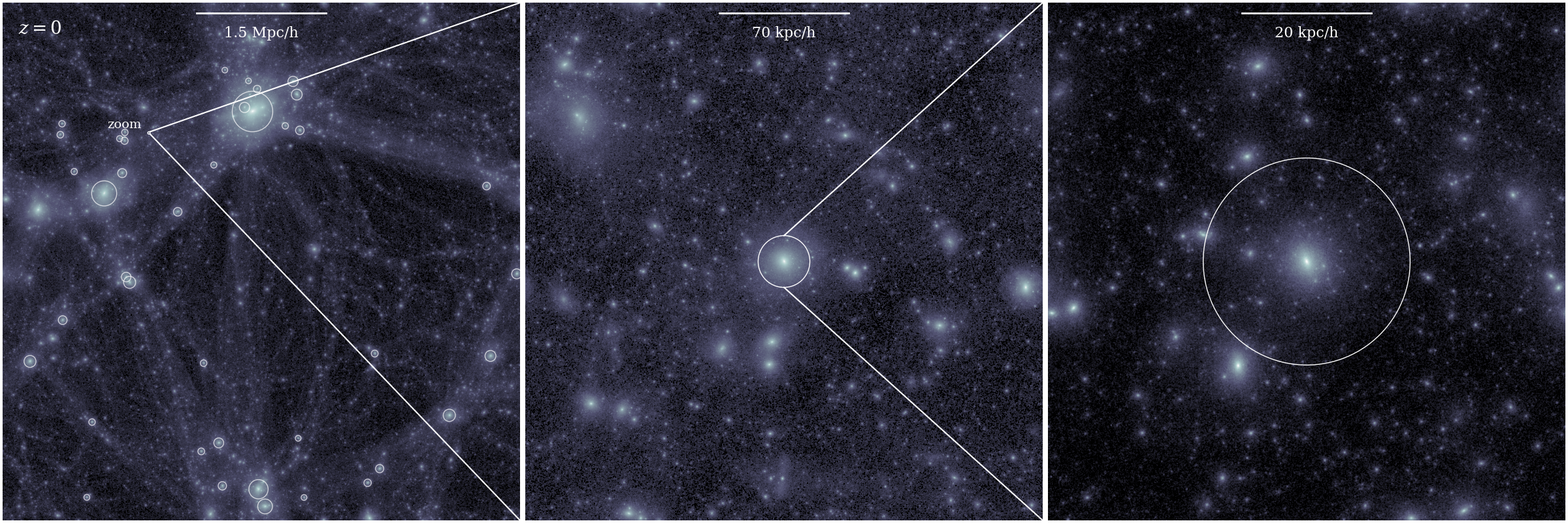}
     \caption{Projected DM distribution at $z = 0$. Three panels show different spatial scales, progressively zooming in from left to right. \texttt{Left}: parent simulation box of the DM-only run, including the target UFD and 40 nearby massive halos, marked as white circles. The most massive halo in the box has a virial mass of $\sim10^{12}\Msun$ at $z=0$. \texttt{Middle}: zoom-in region centered on the target UFD. \texttt{Right}: further zoomed-in region encompassing the target UFD (HALO1-\BG\nspace, see Table \ref{tab:set}), which has a virial radius of $\sim$ 10\,kpc and a virial mass of $\sim2.5\times10^8\Msun$ at $z=0$.}
     \label{fig:dmmorphology}
\end{figure*}

We model cosmic reionization using two schemes: Background (\BG) and Radiation (\Rad). Our goal is to investigate how a more realistic treatment of reionization affects UFD formation and evolution; this requires an RT-based patchy reionization model together with sufficient mass resolution to resolve UFD-scale systems. Because fully coupled RHD box simulations remain computationally expensive, we adopt a simplified framework for UFD zoom-in simulations. Both schemes share the following four steps: (i) we run a low-resolution ($m_{\rm DM}\approx10^{6}\Msun$) DM-only simulation in a parent box of size $L_{\rm box}=6.25h^{-1}\rm{cMpc}$ (illustrated in Fig \ref{fig:dmmorphology}); (ii) we identify 40 massive halos and our target UFD halos ($M_{\rm vir}\sim10^{8}\Msun$ at $z=0$) in the parent box and construct merger trees that follow their main progenitors. These massive halos allow us to quantify the local epoch of reionization (EoR) environment experienced by each target halo; (iii) we estimate the stellar masses of the 40 massive halos as functions of halo mass and redshift using \universemachine\nspace (\citealp{Behroozi2019}); and (iv) we derive their SEDs with \st\nspace (\citealp{Leitherer1999}).
\par
In the third step, we adopt the analytic stellar mass to halo mass (SMHM) relations given by \universemachine\nspace (\citealp{Behroozi2019}) to estimate the stellar masses of the source halos. \universemachine\nspace infers the SMHM relation empirically by populating DM-only halo merger trees and calibrating a galaxy-halo connection model to multiple observables (e.g., stellar mass functions, galaxy clustering, and weak lensing) using Markov chain Monte Carlo (MCMC). Although these fits are primarily constrained at slightly higher halo masses ($M_{\rm halo} \gtrsim 10^{10}\Msun$) than our source-halo range ($10^{9}\Msun \lesssim M_{\rm halo} \lesssim 10^{10}\Msun$), we apply the relation as an approximate extrapolation. 

\par
As a final step, we derive the SED of the source halos using \st\nspace the package (\citealp{Leitherer1999}). \st\nspace generates synthetic galaxy spectra from the stellar masses estimated in the previous step, for which we adopt the Geneva stellar evolution tracks and assume that massive OB stars dominate the production of ionizing photons in the early Universe. Motivated by the $\sim 10\,\rm{Myr}$ main sequence lifetimes of OB stars, we compute the SED in $\Delta t=10\,\rm{Myr}$ bins using the newly formed stellar mass $\Delta M_{\star}$, within each time interval. A key limitation of this approach is that the stellar masses inferred from halo growth are monotonic, because halo masses increase almost continuously even during periods when star formation may be suppressed. Consequently, $\Delta M_{\star}$ remains non-zero at all times in our model, which may lead to an overestimate of the source-halo SEDs compared to hydrodynamic simulations that naturally capture quiescent and bursty star formation.

\par
\begin{figure}
    \centering
    \includegraphics[width=1.0\linewidth]{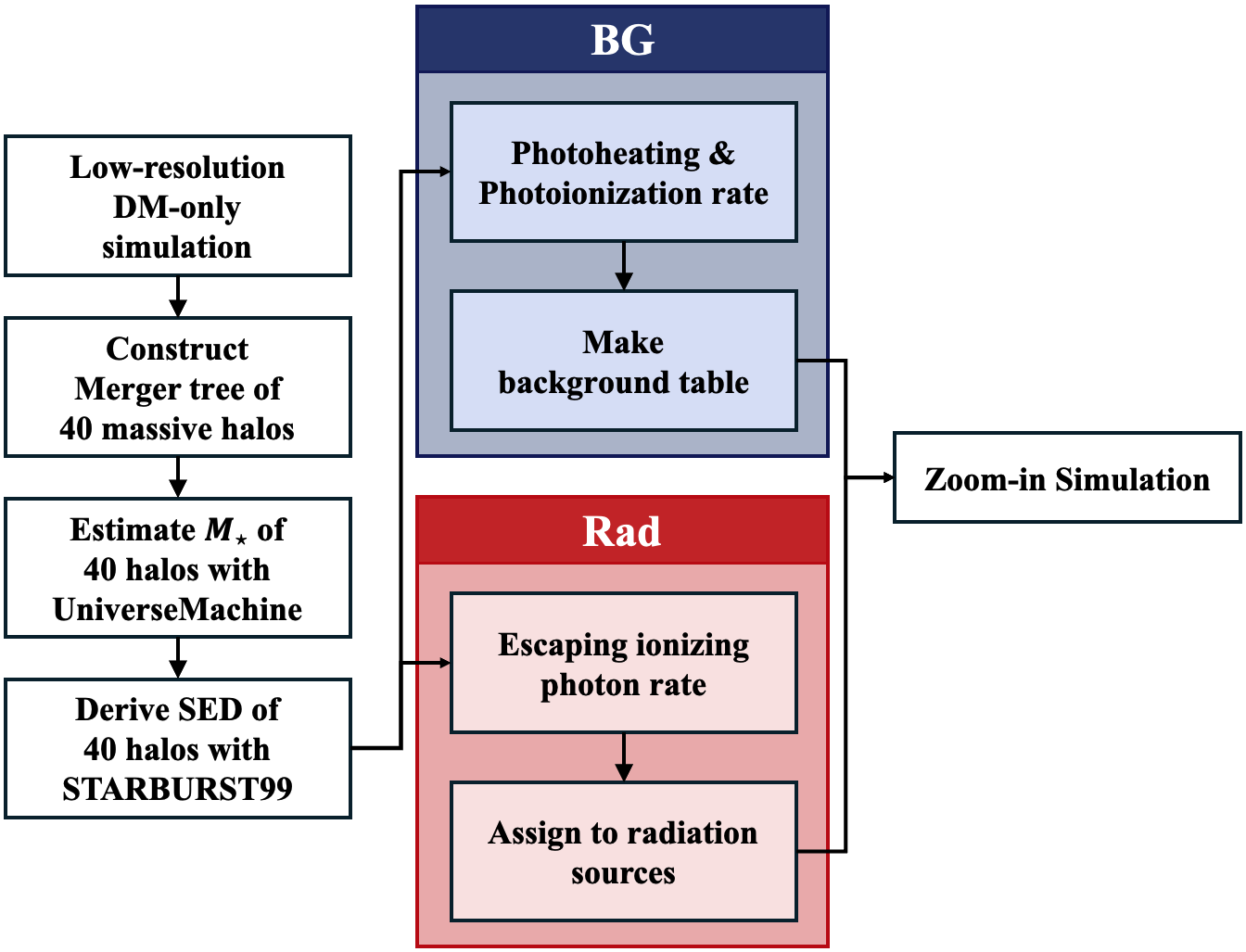}
    \caption{Flowchart illustrating the setup procedure for running simulations with two different reionization schemes. We first run a low-resolution DM-only simulation to construct the merger trees of the target UFD and 40 massive halos in the simulation box. We then use \universemachine\nspace (\citealp{Behroozi2019}) to estimate the stellar mass of each halo as a function of halo mass and redshift. Next, we use the \st\nspace (\citealp{Leitherer1999}) package to derive the SEDs of the 40 halos from their stellar masses. We first multiply the intrinsic SEDs by the redshift-dependent escape fraction of \citet{Faucher_Gigu_re_2020}. \BG: We compute photoheating and photoionization rates from the escaping SEDs, tabulate them, and use these tables as a homogeneous background in the zoom-in simulations. \Rad: We compute the escaping ionizing photon production rate from the same SEDs and assign it to radiation sources in the zoom-in simulations.}
    \label{fig:flowchart}
\end{figure}

Thereafter, we construct the \BG\nspace and \Rad\nspace schemes using the \st\nspace outputs. The two schemes are based on the same underlying outputs, but differ in how these quantities are applied in the simulations. The procedures for modeling the two schemes are briefly summarized in Fig \ref{fig:flowchart}. We follow the evolution of our systems down to $z=5$, by which time hydrogen reionization is observationally inferred to be complete (e.g., \citealp{Becker2015}; \citealp{Mcgreer2015}; \citealp{Bosman2022}). We do not extend the simulations to lower redshift, because the relative importance of massive star-forming galaxies and AGNs to the meta-galactic ionizing background increases toward $z \lesssim 5-3$ (e.g., \citealp{Faucher_Gigu_re_2020}; \citealp{Finkelstein2022}), while our radiation model includes ionizing sources only from star-forming galaxies. Our primary interest is the evolution of UFDs during EoR, and therefore we activate reionization effects on $5 \lesssim z \lesssim 7$, following the convention of our previous simulations (e.g., \citealp{Jeon_2017, Jeon2021b, Kim2023, JeonKo2026}). This redshift interval is identical in the \BG\nspace and \Rad\nspace runs, and we examine the physical implications of this choice in Section \ref{caveat}.

\par
Both schemes take as input the SEDs after an ionizing photon escape fraction, $f_{\rm esc}$, has been applied. It is defined as the fraction of ionizing photons produced by star-forming galaxies that escape into the IGM. Although $f_{\rm esc}$ likely depends on galaxy mass, environment, and the burstiness of star formation, for simplicity, we adopt the redshift-dependent analytic prescription given by Eq. (12) of \citealp{Faucher_Gigu_re_2020}, which yields $f_{\rm esc} \simeq 0.03-0.05$ over $5 \le z \le 7$, and apply it identically in the \BG\nspace and \Rad\nspace schemes. All photoionization and photoheating rates (\BG) and photon production rates (\Rad) below are computed from these escaping SEDs.

\subsubsection{Scheme 1: Background (\rm{\BG})}
The first approach is the uniform reionization background, which contains the redshift-dependent photoionization and photoheating rate for $\rm{H}$ $\rm{I}$, $\rm{He}$ $\rm{I}$, and $\rm{He}$ $\rm{II}$. The photoionization rate ($\Gamma_{\rm i}$) and photoheating rate ($\mathcal{H}_{\rm i}$) are calculated from equations below (\citealp{OsterbrockFerland2006}).

\begin{equation}
    \Gamma_{\rm {i}} = \int^{\infty}_{\nu_{\rm min}} \frac{F_{\nu} \sigma_{\nu}}{h\nu} d\nu
\end{equation}

\begin{equation}
    \mathcal{H}_{\rm {i}} = \int^{\infty}_{\nu_{\rm min}} F_{\nu}\sigma_{\nu} \Big( 1-\frac{\nu_{\rm min}}{\nu} \Big) d\nu
\end{equation}

\par
\begin{figure}
    \centering
    \includegraphics[width=1.0\linewidth]{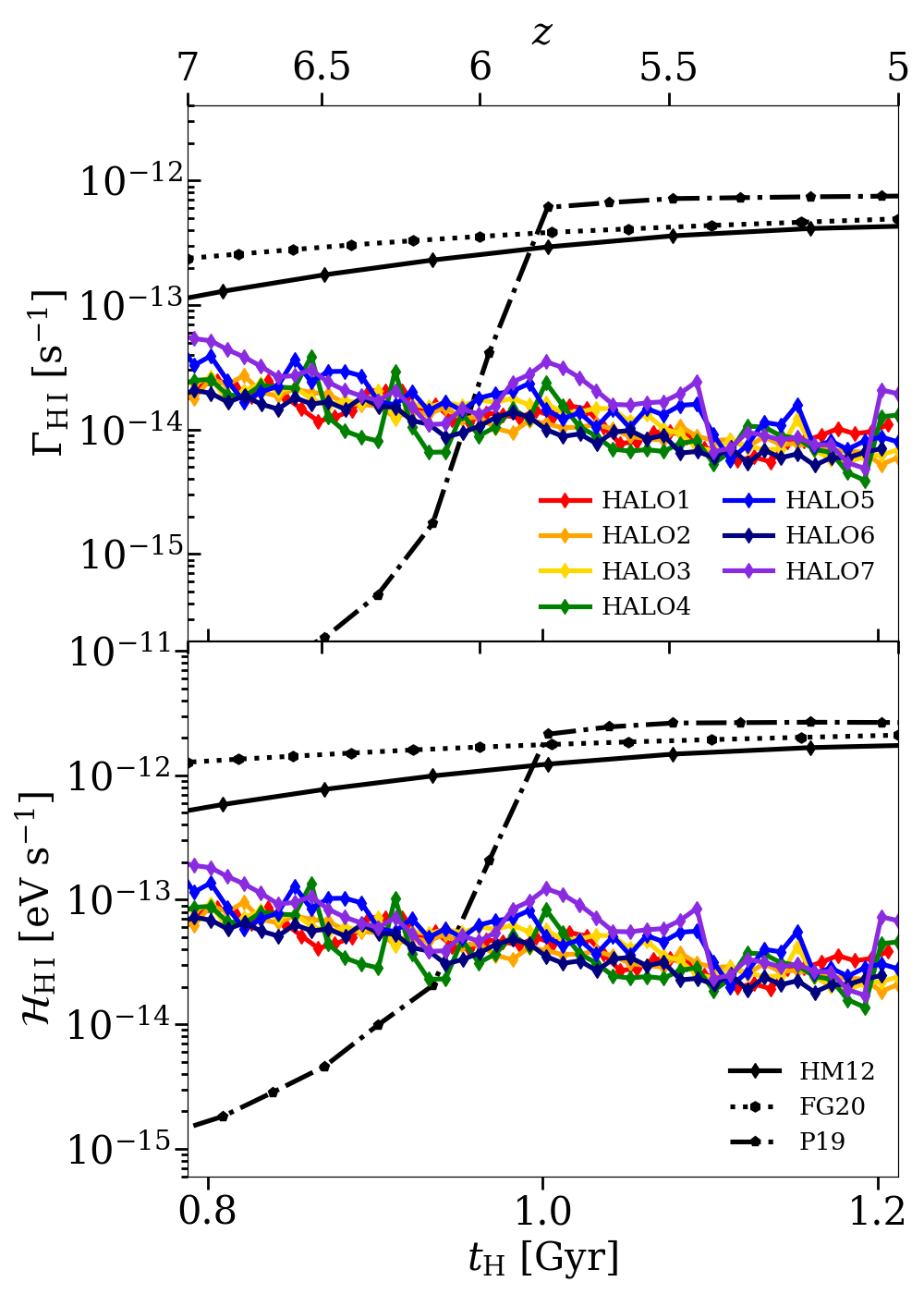}
    \caption{Photoionization (upper panel) and photoheating (lower panel) rates of hydrogen at the position of each target UFD, computed from the SEDs of the 40 massive galaxies after the \citetalias{Faucher_Gigu_re_2020} escape fraction ($f_{\rm esc} \simeq 0.03-0.05$) has been applied, from $z=7$ to $z=5$. In the \BG\nspace runs, we tabulate these rates and use them as a uniform background to model the effect of reionization. For comparison, three UVB tables are shown as black lines; the solid, dotted, and dash-dotted lines correspond to \citealp{HaardtMadau2012} (HM12), \citealp{Faucher_Gigu_re_2020} (FG20), and \citealp{Puchwein_2019} (P19), respectively. At the onset of reionization at $z=7$, the \BG\nspace rates lie a factor of $\sim5$ below \citetalias{HaardtMadau2012} and $\sim 10$ below \citetalias{Faucher_Gigu_re_2020}. Because our rates decline mildly while the tabulated UVBs continue to rise, the gap widens to roughly 1.5 dex by $z=5$.}
    \label{fig:UVB}
\end{figure}

In these equations, $F_{\nu}$ denotes the escaping ionizing flux from the massive halos incident on the target UFD, computed from the SEDs described above. The ionization cross-section, $\sigma_{\nu}$, quantifies the likelihood that an atom or molecule absorbs a photon of frequency $\nu$ and becomes ionized  (\citealp{OsterbrockFerland2006}). The threshold frequency for ionization is denoted by $\nu_{\rm min}$, corresponding to $h\nu_{\rm min} = \rm{13.6\,eV}$, $h\nu_{\rm min} = \rm{24.6\,eV}$, and $h\nu_{\rm min} = \rm{54.4\,eV}$ for $\rm{H\,I}$, $\rm{He\,I}$, and $\rm{He\,II}$, respectively. After computing the photoionization and photoheating rates contributed by the SEDs of 40 surrounding massive halos to a given target UFD, we sum these contributions and tabulate a single, target-specific UVB for each UFD. The resulting photoionization and photoheating rates are displayed in Fig \ref{fig:UVB}.

\par
Such reionization effects are implemented by switching on the UVB and applying the resulting photoionization and photoheating rates to all gas particles in the simulation box irrespective of position. These rates are coupled to the non-equilibrium chemistry network to evolve the H/He ionization states, while self-shielding in dense gas is modeled by attenuating the UVB as $\exp (-N_{\rm HI} \bar{\sigma}_{\rm ion})$, where $N_{\rm HI}=xn_{\rm HI}$ ($x$ is the SPH kernel size, $n_{\rm HI}$ the neutral hydrogen number density, and $\bar{\sigma}_{\rm ion}$ the frequency-averaged $\rm {H\,I}$ photoionization cross-section). Although the UVB for \BG\nspace itself encodes an environment-dependent radiation field, reionization is still imposed in a flash-like UVB on/off manner, so the time-dependent propagation of ionization fronts is not explicitly followed.

\subsubsection{Scheme 2: Radiation (\rm{\Rad})}
\label{Radscheme}
\begin{figure}
    \centering
    \includegraphics[width=1.0\linewidth]{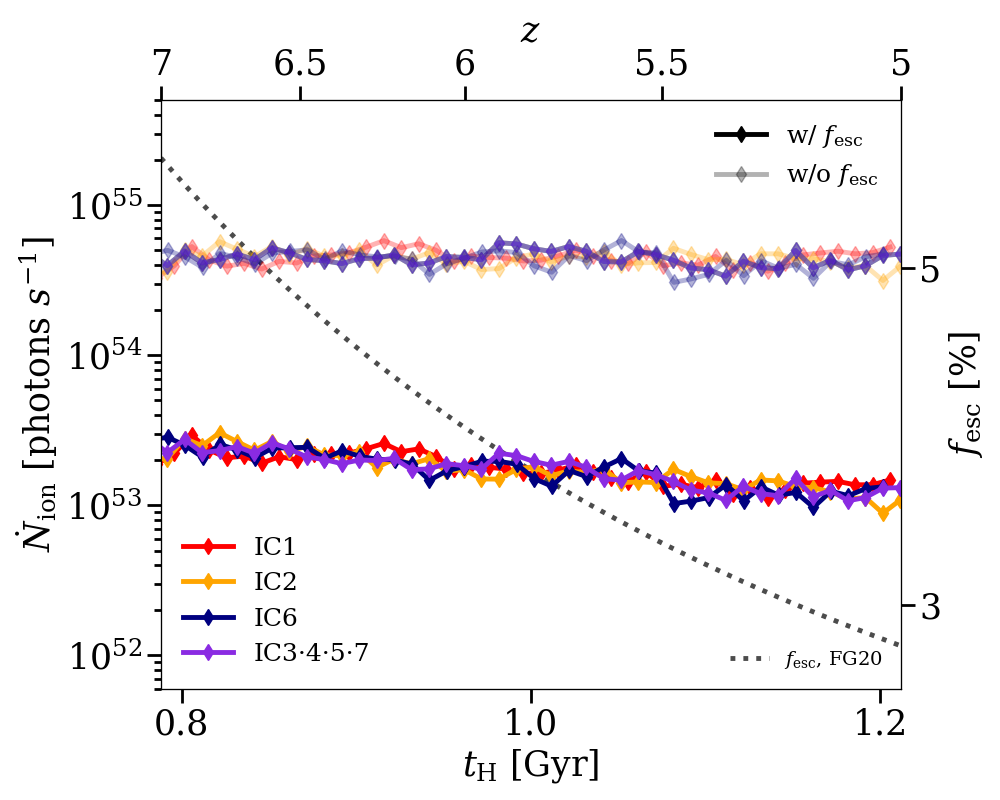}
    \caption{Ionizing photon production rate from $z=7$ to $z=5$, obtained by summing the contributions of 40 massive galaxies. Each color represents the photon production rate associated with the initial condition (IC) in which the target halo resides. Because HALO3, HALO4, HALO5, and HALO7 are generated from the same IC, we utilize the same photon production rate for those halos. We adopt the analytic redshift-dependent escape fraction ($f_{\rm{esc}}$) from \citealp{Faucher_Gigu_re_2020} (FG20), shown as the dotted line on the secondary $y$-axis. The intrinsic photon production rate before applying the escape fraction is labeled as ``w/o $f_{\rm esc}$'' and shown as semi-transparent lines, while the escaping photon production rate after applying this escape fraction is labeled as ``w/ $f_{\rm esc}$'' and shown as opaque lines. Both schemes, \BG\nspace and \Rad, use the escaping quantities; the \Rad\nspace runs assign the ``w/ $f_{\rm esc}$'' rates shown here directly to the radiation sources.}
    \label{fig:nion}
\end{figure}

The second approach is a source-driven, RT-based patchy reionization model. In the first scheme, \BG, cosmic reionization is approximated by applying a pre-tabulated, spatially uniform UVB to the hydrodynamic simulation. In this scheme, instead of imposing a UVB, we introduce discrete radiation sources that represent massive halos in the low-resolution simulation and propagate their ionizing photons through the RT module, \traphic. Using \st, which provides the ionizing photon production rates (photons s$^{-1}$) for $\rm {H\,I}$-, $\rm {He\,I}$-, and $\rm {He\,II}$-ionizing radiation, we obtain the intrinsic ionizing photon production rate, $N_{\rm ion}$, of each massive halo. Applying the escape fraction introduced in Section \ref{CR} gives the escaping ionizing photon rate, $\dot{N}_{\rm ion,\,esc} \equiv f_{\rm esc} \dot{N}_{\rm ion}$, which we assign to each radiation source. Figure \ref{fig:nion} shows the total escaping ionizing photon rate obtained by summing the contributions of the 40 radiation sources for each initial condition.

\par
\begin{figure}
    \centering
    \includegraphics[width=0.95\linewidth]{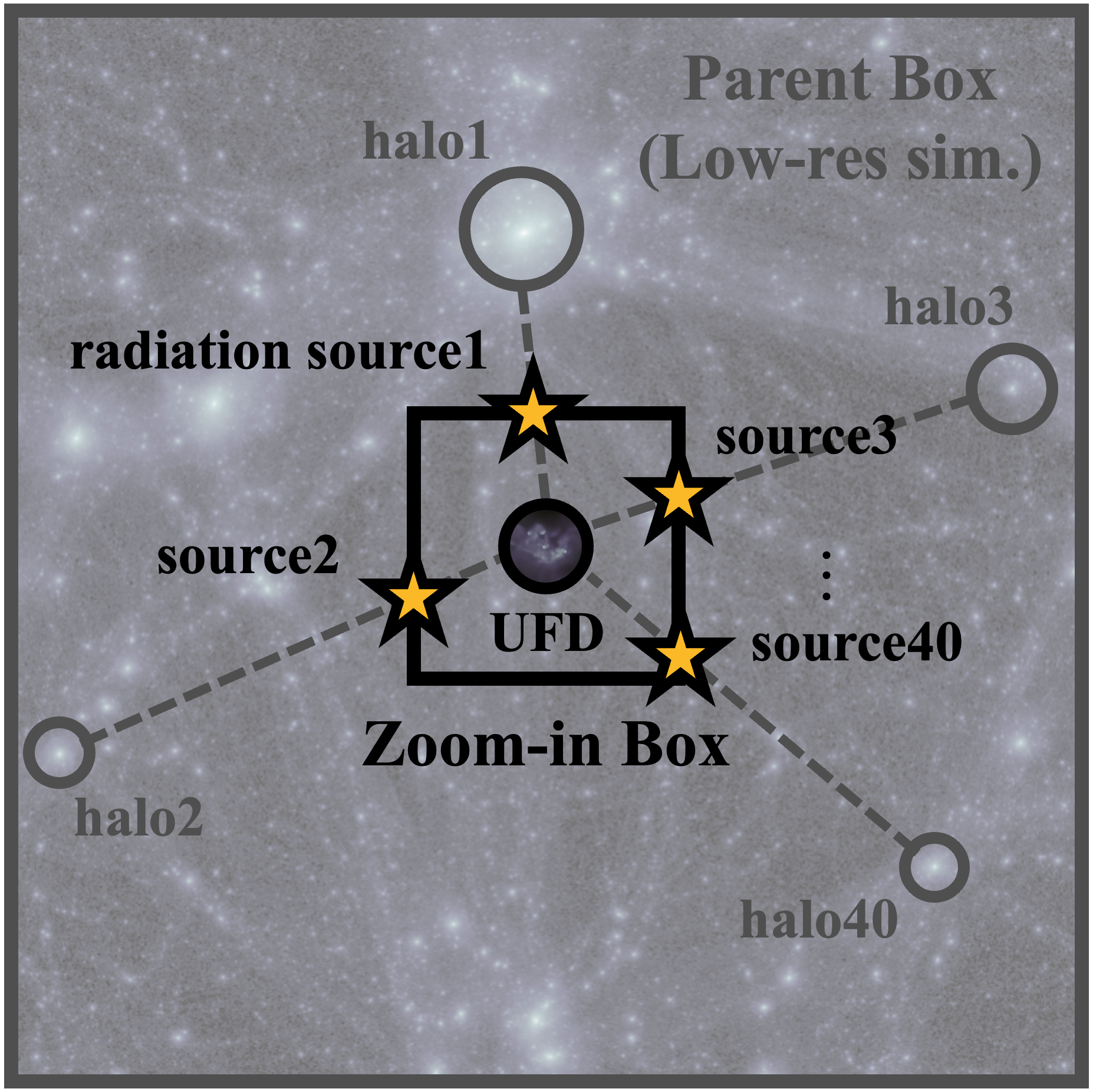}
    \caption{Schematic illustration of how 40 radiation sources each representing a massive halo in the low-resolution parent box, are placed in the zoom-in region. Because the RT calculations are performed only within the high-resolution zoom-in volume, we cannot place radiation sources directly at the positions of the massive halos identified in the parent low-resolution simulation. Instead, we use those halo positions as references and place the sources within the zoom-in region accordingly. The gray circles represent the massive halos in the parent box, and the black rectangle at the center marks the zoom-in region. For each massive halo, we place a radiation source (yellow star symbol) at the intersection between the zoom-in box and the line connecting the massive halo to the target UFD at the center of the zoom-in region.}
    \label{fig:illust}
\end{figure}
\begin{figure}
    \centering
    \includegraphics[width=1.0\linewidth]{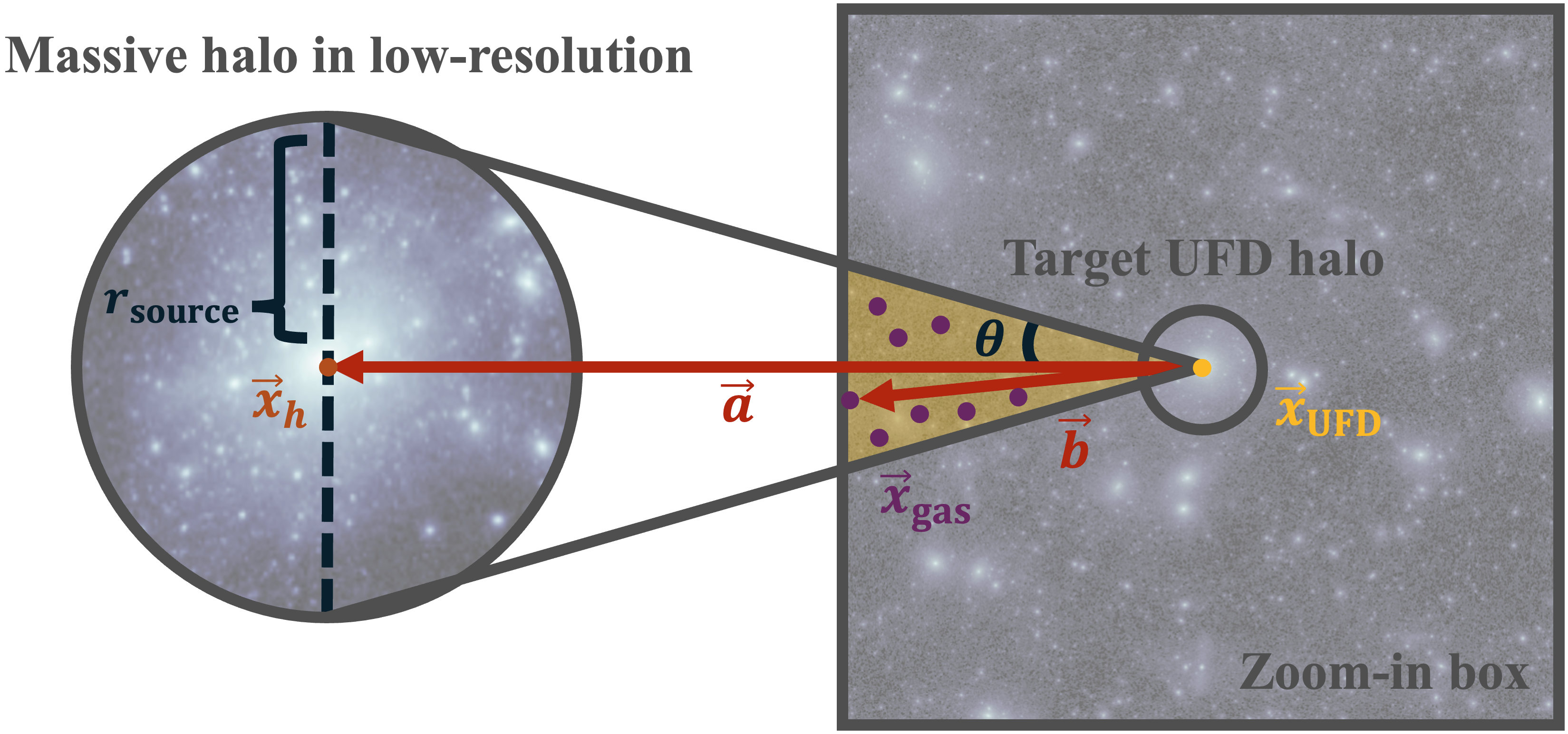}
    \caption{Illustration of the procedure used to place a radiation source representing a massive halo (Steps~1--4 in Section~\ref{Radscheme}). We define $\vec{a} \equiv \vec{x}_{h} - \vec{x}_{\rm UFD}$ and $\vec{b} \equiv \vec{x}_{\rm gas} - \vec{x}_{\rm UFD}$, and the opening angle $\theta = \arccos ( (\vec{a} \cdot \vec{b}) \, / \, (|\vec{a}||\vec{b}|))$. Gas particles satisfying $\theta < \arctan (r_{\rm source}\,/\,|\vec{a}|)$ lie inside the yellow-shaded cone. Here, we set the effective radius of massive galaxy with virial radius of halo, $r_{\rm source} = 0.01 R_{\rm vir}$ (e.g., \citealp{Kravtsov2013}). Among high-resolution gas particles (purple circles) within the cone, we select the one with the largest $|\vec{b}|$ (i.e., the farthest from the UFD center) and adopt its position as the radiation source location representing the massive halo.
    }
    \label{fig:illust2}
\end{figure}

With the escaping ionizing photon rate of each radiation source in hand, the remaining question is how to assign their locations. Since these rates are derived for massive halos identified in the parent low-resolution DM-only simulation, we also use their halo positions to place the radiation sources, as briefly illustrated in Fig. \ref{fig:illust}. Because photoionization and photoheating by photon packets are applied only within the zoom-in region (and not in the padding or low-resolution region), all sources must be placed inside the zoom-in volume to ensure that their radiative impact is properly captured. In addition, if a source is located too close to the target UFD, the resulting ionization front can traverse the region on a timescale shorter than the snapshot spacing, effectively producing a nearly instantaneous, flash-like manner of UVB implementation rather than a gradually propagating front. To avoid this behavior while satisfying the zoom-in constraint, we place each radiation source at the farthest possible position from the UFD center within the zoom-in region. This setup enables us to model the radiative impact of reionization on the target UFD in an outside-in manner. The detailed procedure is shown below and illustrated in Fig \ref{fig:illust2}.

\begin{enumerate}
    \item \textbf{Define the direction of the source.}\\
    For each massive halo, we define the vector \[\vec{a} \equiv \vec{x}_{h} - \vec{x}_{\rm UFD},\] where $\vec{x}_{h}$ and $\vec{x}_{\rm UFD}$ are the centers of the massive halo and the target UFD halo, respectively, both taken from the low-resolution DM-only simulation.

    \item \textbf{Identify candidate gas positions.}\\
    For each gas particle within the zoom-in volume, we define \[\vec{b} \equiv \vec{x}_{\rm gas} - \vec{x}_{\rm UFD},\] where $\vec{x}_{\rm gas}$ is the position of the gas particle.

    \item \textbf{Select the farthest position within the cone.}\\
    We compute the opening angle $\theta$ between $\vec{a}$ and $\vec{b}$ as \[\cos \theta = \frac{\vec{a} \cdot \vec{b}}{|\vec{a}| |\vec{b}|}.\] We then select, among all gas particles satisfying \[\theta < \arctan \Big( \frac{r_{\rm source}}{|\vec{a}|}\Big),\] the one with the longest distance $|\vec{b}|$. The position of this gas particle is assigned as the radiation source location. Here we assume the effective radius of the massive galaxy $r_{\rm source}=0.01R_{\rm vir}$ (e.g., \citealp{Kravtsov2013}), where $R_{\rm vir}$ is the virial radius of the corresponding massive halo.

    \item \textbf{Update the source position at each time step.}\\
    We repeat this procedure (1--3) at each simulation time step.
\end{enumerate}

\par
Because each source is placed at the zoom-in boundary rather than at the true position of the massive halo it represents, we rescale its escaping photon rate by $(|\vec{b}|/|\vec{a}|)^{2}$ so that the ionizing flux received at the target UFD matches that of a source at its true distance $|\vec{a}|$. Each radiation source is then assigned this renormalized rate, $\dot{N}_{\rm ion,\,esc} \times (|\vec{b}|/|\vec{a}|)^{2}$, and \traphic\nspace propagates the corresponding photon packets, from which the photoionization and photoheating of each gas particle are computed self-consistently.

\par
Over $5 \leq z \leq 7$, the source subtends an angular diameter of $\theta \sim 0.1-0.2^{\circ}$, corresponding to $r_{\rm source} \,/\, |\vec{a}| \sim 10^{-3}$. Since $r_{\rm source} \ll |\vec{a}|$ for all source-target pairs, the finite angular extent of the source has a negligible impact on the radiative flux at the target, and we therefore model each massive galaxy as a point source, following the standard treatment in cosmological RT simulations (e.g., \citealp{iliev2006}).

\par
Two schemes introduced in this study, \BG\nspace and \Rad, rely on the same \st\nspace outputs and the same escape fraction, both derived from the same parent low-resolution simulation. Although both prescriptions incorporate the local environment by considering massive halos identified in the parent run, there is a key distinction in how the reionization effects are applied to the target UFD. \BG\nspace imposes a flash-like spatially uniform irradiation once the background is turned on, while \Rad\nspace follows the propagation of ionizing radiation with RT, providing a more physically motivated implementation.

\begin{table*}
\centering
\setlength{\tabcolsep}{6.5pt}
\renewcommand{\arraystretch}{1.0}
\begin{tabular}{c c c c c c c c}
\hline
\hline
Name & ${M_{\rm vir, z=0}} \ [10^{8} \Msun]$ & ${M_{\rm vir, z=5}} \ [10^{8} \Msun]$ & ${M_{\rm gas} \ [10^{5} {\Msun}]}$ & $M_{\rm \star} \ [10^{3} {\Msun}]$ & $\langle[{\rm Fe/H}]\rangle$ & $z_{\rm{\tau_{90}}}$ & Reionization approach \cr
\hline
\hline 
{\sc H1-BG} & \multirow{2}{*}{2.46} & \multirow{2}{*}{0.238}& {1.47} & {0.679} & {-3.06} & {10.5} & Background\cr
{\sc H1-Rad} & & & {10.3} & {7.52} & {-4.09} & {5.04} & Radiation\cr
\hline
{\sc H2-BG} & \multirow{2}{*}{3.28} & \multirow{2}{*}{0.278}& {2.17} & {2.14} & {-2.72} & {7.82} & Background\cr
{\sc H2-Rad} & & & {31.9} & {27.0} & {-2.50} & {5.17} & Radiation\cr
\hline
{\sc H3-BG} & \multirow{2}{*}{4.71} & \multirow{2}{*}{0.309}& {1.97} & {0.00} & {-} & {9.16} & Background\cr
{\sc H3-Rad} & & & {3.83} & {32.0} & {-2.69} & {5.21} & Radiation\cr
\hline
{\sc H4-BG} & \multirow{2}{*}{3.74} & \multirow{2}{*}{0.409}& {3.03} & {0.116} & {-5.76} & {9.97} & Background\cr
{\sc H4-Rad} & & & {5.20} & {0.120} & {-5.76} & {6.98} & Radiation\cr
\hline
{\sc H5-BG} & \multirow{2}{*}{9.77} & \multirow{2}{*}{0.561}& {23.39} & {3.11} & {-3.03} & {7.40} & Background\cr
{\sc H5-Rad} & & & {49.2} & {38.6} & {-2.71} & {5.17} & Radiation\cr
\hline
{\sc H6-BG} & \multirow{2}{*}{5.90} & \multirow{2}{*}{0.574}& {33.61} & {2.32} & {-3.14} & {8.85} & Background\cr
{\sc H6-Rad} & & & {74.1} & {2.45} & {-3.08} & {5.33} & Radiation\cr
\hline
{\sc H7-BG} & \multirow{2}{*}{8.58} & \multirow{2}{*}{1.95}& {23.90} & {33.8} & {-2.87} & {7.33} & Background\cr
{\sc H7-Rad} & & & {20.4} & {33.9} & {-2.87} & {7.33} & Radiation\cr
\hline
\hline
\end{tabular}
\caption{Summary of the simulation sets. Column (1): Name of run. Column (2): Halo mass of the target UFD at $z=0$. Column (3): Halo mass of the target UFD at $z=5$. Column (4): Gas mass of the target UFD at $z=5$. Column (5): Stellar mass of the target UFD at $z=5$. Column (6): Averaged stellar iron-to-hydrogen ratio of the target UFD at $z=5$. Column (7): The time when 90\% of the total stellar mass has formed. Column (8): The way we incorporate the reionization effect to each simulation set, background (\BG) and radiation (\Rad).}
\label{tab:set}
\end{table*}

\section{Results}
\label{Sec:Result}

In this section, we present the results of our simulations for seven UFDs with two reionization prescriptions: \BG\nspace (flash-like) and \Rad\nspace (RT-based). The properties of these simulation sets are summarized in Table \ref{tab:set}. In Section \ref{KDZ}, we describe differences in the gas properties within the zoom-in region under the two models. In Section \ref{KDU}, we examine how the star formation activity in the target UFD varies between the two prescriptions. In Section \ref{EVOL}, we analyze the diverse evolutionary pathways of UFDs in the \Rad\nspace scheme. Finally, in Section \ref{OBSERVABLE}, we analyze some observable quantities.

\par
Unless otherwise noted, we adopt the following consistent plotting conventions throughout the figures. Each halo is shown with a fixed color, and we consistently adopt this color scheme across all panels. The \BG\nspace runs are plotted with solid lines, while the \Rad\nspace runs are plotted with dash-dotted lines. In time-evolution figures, the vertical shaded region indicates the redshift range of reionization adopted in our simulations $5 \lesssim z \lesssim 7$. The halos are labeled in ascending order of their virial mass at $z=5$, such that HALO1 represents the least massive system and HALO7 denotes the most massive at this epoch.

\subsection{Key differences in the zoom-in volume}
\label{KDZ}
\subsubsection{Gas properties}

\begin{figure*}
    \centering
    \includegraphics[width = 175mm]{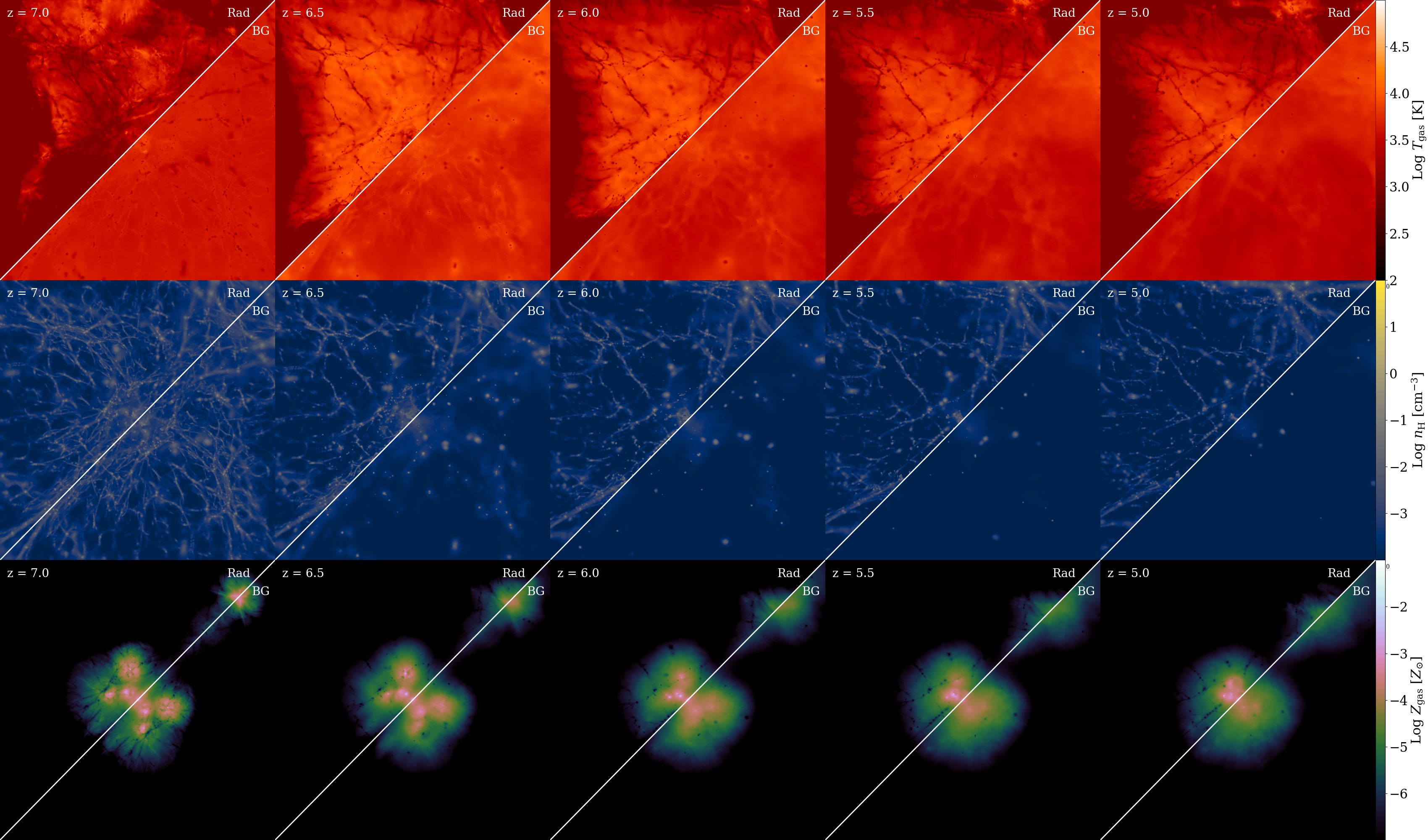}
    \caption{Gas properties within the zoom-in box of HALO2. From left to right, the columns show snapshots from $z=7$ to $z=5$. Each row presents projections of gas temperature, hydrogen number density, and gas-phase metallicity, respectively. Each panel is displayed in a mirror-symmetric layout with respect to the diagonal, with the upper-left half showing \Rad\nspace and the lower-right half showing \BG. In the hydrogen density maps, \BG\nspace loses most filamentary structure by $z\approx 6.0$, while \Rad\nspace still retains dense clumps even at $z=5$. In the temperature distribution, gas in \BG\nspace undergoes global heating from $z=7$, becoming low-density ($\sim10^{-3}\rm{cm}^{-3}$) and hot ($\sim10^4\rm{K}$) by $z=5$, whereas \Rad\nspace preserves cold gas ($\sim10^3\rm{K}$) along dense structures until $z=5$. In the metallicity maps, metals in \BG\nspace diffuse more rapidly, leaving no high-metallicity regions compared to \Rad. Global heating in \BG\nspace accelerates the disruption of dense structures, leading to the absence of cold, dense gas required for star formation.}
    \label{fig:gasproperties}
\end{figure*}

The primary signatures of cosmic reionization are manifested in the gas properties. To illustrate the impact of reionization on the gas phase, Figure \ref{fig:gasproperties} presents projected maps of gas properties within the HALO2 zoom-in region. These maps encompass the entire high-resolution Lagrangian volume with a comoving side length of $\sim 300\,\rm{ckpc}$, projected along the z-axis to capture the global structure of the target system and its surrounding environment. From top to bottom, we present gas temperature, hydrogen number density, and gas-phase metallicity. Each row consists of five snapshots ordered from left to right, covering $z=7$ to $z=5$, during which reionization is active in our simulations. For a direct visual comparison between the two reionization prescriptions, each panel is split by the white diagonal: the upper-left shows the \Rad\nspace run, while the lower-right shows the \BG\nspace run, forming a mirror-symmetric layout.

\par
In the temperature maps, \BG\nspace exhibits a rapid, nearly global heating to the canonical photoheated IGM temperature of $\sim 10^{4}\, \rm{K}$ when reionization becomes active at $z=7$, producing a mean gas temperature of $\langle\, T_{\rm gas,\, BG}\, \rangle \approx 10^{3.7} \rm {K}$. In contrast, \Rad\nspace shows a more gradual photoheating process as the ionizing radiation propagates through the zoom-in region, leaving the gas relatively cool at $z=7$ with $\langle \, T_{\rm gas, \, Rad} \, \rangle \approx 10^{2.9} \, \rm{K}$. Correspondingly, the majority of dense hydrogen clumps (defined as gas with $n_{\rm H}>10 \, \rm{cm}^{-3}$) are no longer present by $z \approx 6.0$ in \BG, whereas \Rad\nspace retains such dense clumps even in $z=5$, displayed in the second row. As seen in the bottom row, the gas-phase metallicity appears more diffusely distributed in the \BG\nspace run compared to the more localized enrichment patterns observed in the \Rad\nspace case.

\par
As described in Section \ref{starformation}, star formation in our simulations occurs only in gas above the threshold density $n_{\rm H,\, thr} = 10^{2}\, \rm{cm}^{-3}$. Because the majority of dense clumps are dissipated by $z \approx 6.0$ in \BG, star formation is suppressed strongly thereafter. This suppression is also reflected in the number of CCSNe from Pop II-indiv stars: between $z=7$ and $z=6.5$, we count 62 CCSN events in \Rad\nspace but only 6 in \BG\nspace (a factor of $\sim 10$ lower). Consequently, while initial photoheating associated with reionization brings gas temperatures closer to the canonical scale $\sim 10^{4}\, {\rm K}$ in \BG, the more frequent SN feedback in \Rad\nspace produces additional high-temperature gas, raising both the mean and maximum temperature in the zoom-in region. As a result, despite the flash-like heating at $z=7$ in \BG, the subsequent snapshots ($z \lesssim 6.5$) show higher temperatures in \Rad, with $\langle\, T_{\rm gas} \,\rangle \sim 10^{3.9}\, \rm{K}$ compared to $\sim 10^{3.6}\, \rm{K}$ in \BG, and with $T_{\rm max}$ reaching $\sim 10^{7}\, \rm{K}$ in \BG, while \Rad\nspace attains even higher values of order $\sim 10^{8}-10^{9}\, \rm{K}$. Overall, \BG\nspace quickly loses dense star-forming gas after the onset of reionization, while \Rad\nspace retains shielded dense clumps capable of forming stars in the zoom-in region.

\begin{figure}
    \centering
    \includegraphics[width=1.0\linewidth]{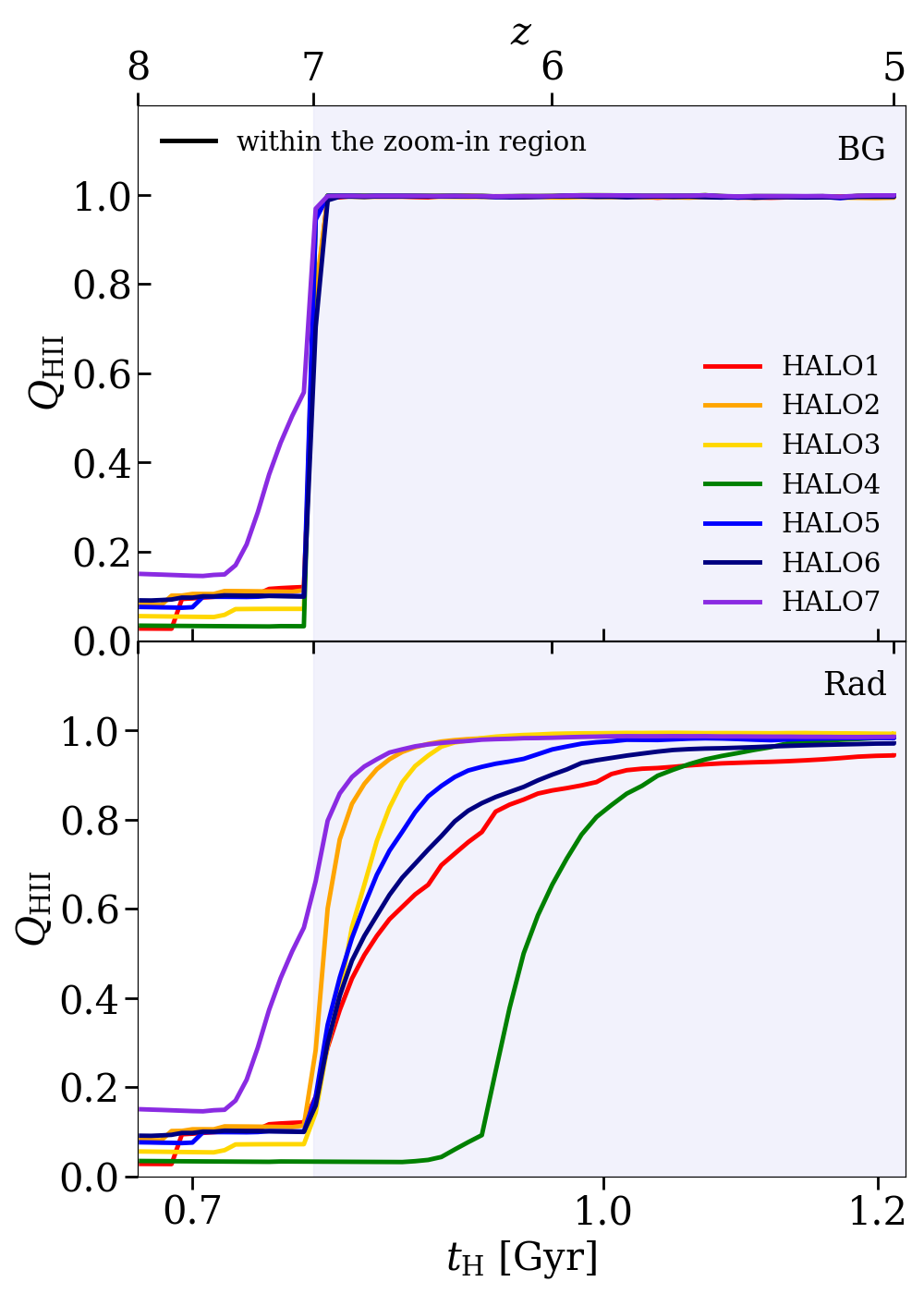}
    \caption{Evolution of the volume filling fraction of ionized hydrogen, $Q_{\rm{HII}}$, within each zoom-in region. The upper panel corresponds to \BG\nspace and the lower to \Rad. For \BG, ionized fractions in all zoom-in regions rise discontinuously to 100\% at the onset of reionization ($z=7$), reflecting the homogeneous application of the reionization field. For \Rad, in contrast, ionization proceeds more gradually, consistent with the propagation of ionizing photons.}
    \label{fig:volumefillingfraction}
\end{figure}

\subsubsection{Volume filling fraction}
We track the gas ionization fraction within the zoom-in volume to check whether the local radiation environment evolves consistently with the broader cosmic reionization history. Figure \ref{fig:volumefillingfraction} shows the volume filling fraction of ionized hydrogen, $Q_{\rm HII}$, as a function of redshift. Here, $Q_{\rm HII}$ is defined as the volume-averaged ionized hydrogen fraction, $Q_{\rm HII} \equiv \sum_{i} x_{{\rm HII},\, i} V_{i} \, / \, \sum_{i} V_{i}$ where $V_{i} = m_{i}\,/\, \rho_{i}$ is the effective volume of each gas particle. We compute $Q_{\rm HII}$ over the entire zoom-in volume. The upper panel presents the \BG\nspace runs, while the lower panel presents the \Rad\nspace runs.

\par
The most noticeable feature is the discontinuous increase in the ionized hydrogen fraction in the zoom-in region under the \BG\nspace scheme, reaching $Q_{\rm HII} \simeq 1$ immediately after the onset of reionization at $z=7$. This behavior is common to all seven zoom-in regions because the \BG\nspace runs adopt similar photoionization and photoheating rates across these regions, as shown in Figure \ref{fig:UVB}.

\par
In contrast, under the \Rad\nspace scheme, the evolution of $Q_{\rm HII}$ is more gradual, with noticeable scatter among the zoom-in regions. The ionized fractions of the zoom-in regions increase over an extended period after $z=7$, reflecting delayed and spatially inhomogeneous reionization as the radiation propagates through the volume. These regions around seven isolated UFDs reach $Q_{\rm HII} \gtrsim 0.95$ by $z \sim 5$, consistent with the reionization constraints from other observations (e.g., \citealp{Becker2015}; \citealp{Mcgreer2015}; \citealp{Bosman2022}). Overall, the comparison highlights that \BG\nspace produces an immediate transition to $Q_{\rm HII} \simeq 1$ in the zoom-in region, while \Rad\nspace produces a delayed and more gradual evolution. We next examine how these scheme-dependent differences in the zoom-in gas translate into the evolution and observable properties of the target UFDs.

\subsection{Key differences in the target UFD}
\label{KDU}

\begin{figure}
    \centering
    \includegraphics[width=1.0\linewidth]{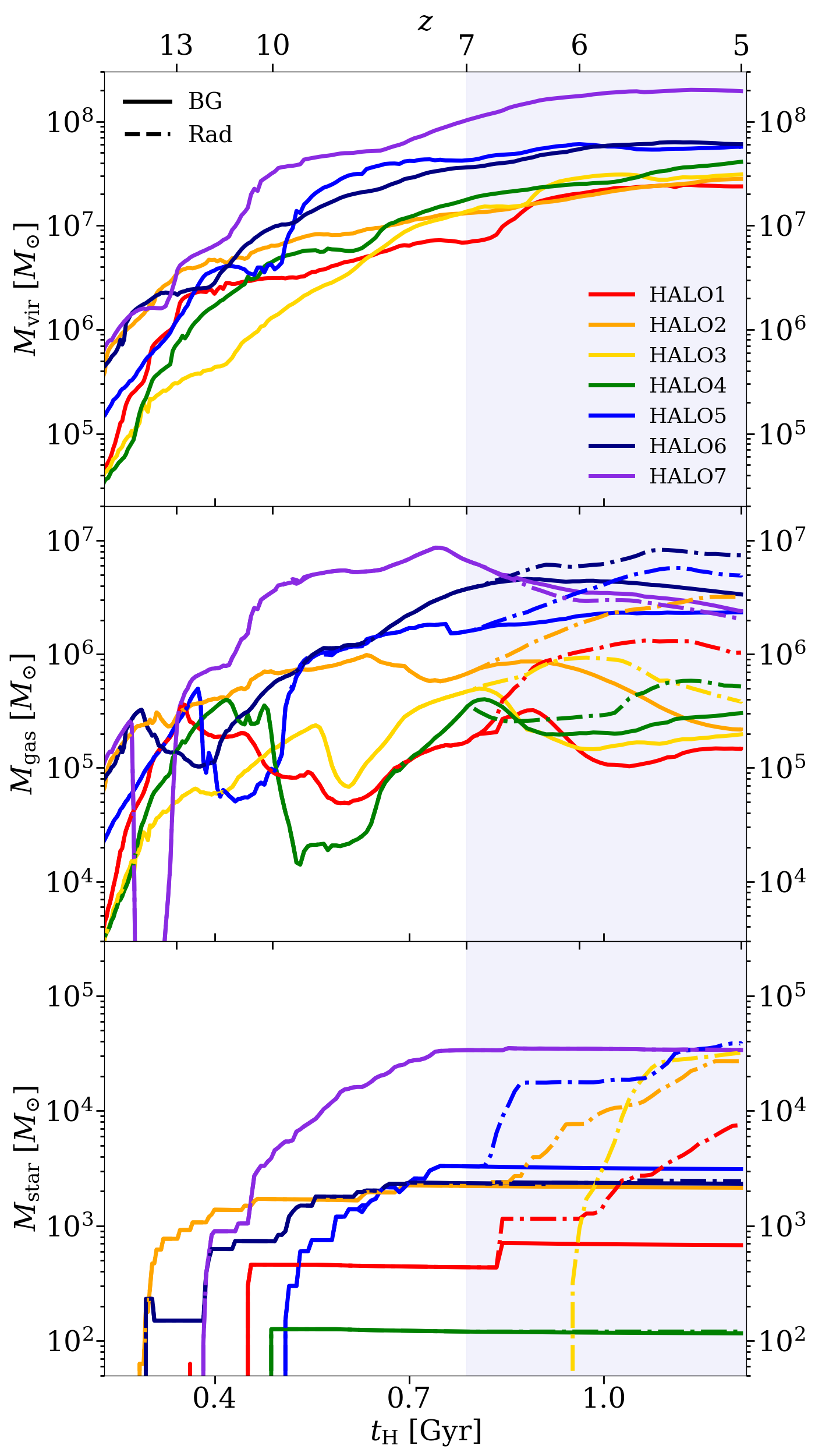}
    \caption{Mass assembly histories of the simulation sets. From top to bottom, the panels show the evolution of virial mass, gas mass, and stellar mass. With the exception of HALO7, all halos retain more gas in the \Rad\nspace runs, ranging from roughly a factor of $\approx2$ (HALO3 and HALO4) up to $\approx15$ (HALO2). A larger gas mass in \Rad\nspace compared to \BG\nspace does not necessarily result in a higher stellar mass: for HALO4 and HALO6, the two schemes yield similar stellar masses.}
    \label{fig:massevolution}
\end{figure}

To characterize the response of the target UFDs to \BG\nspace and \Rad, we first examine their global mass evolution. Figure \ref{fig:massevolution} presents the time evolution of $M_{\rm vir}$, $M_{\rm gas}$, and $M_{\star}$ for our seven halos, with the three quantities shown in the top, middle, and bottom panels, respectively. Because both reionization prescriptions are applied to re-simulations of the same halos, the virial mass growth histories show no systematic differences between \BG\nspace and \Rad. We therefore interpret the differences in $M_{\rm gas}$ and $M_{\star}$ primarily as baryonic responses to reionization modeling.

\par
The evolution of the gas mass exhibits a clear separation between the two schemes for most UFDs. Excluding HALO7, all halos retain more gas in \Rad\nspace than in \BG, with typical enhancements ranging from a factor of $\sim 2$ (HALO3, HALO4) up to $\sim 15$ (HALO2). This difference arises because flash-like global heating in \BG\nspace increases the thermal energy of the gas, making it more difficult for the halo to retain and re-accrete baryons (e.g. \citealp{Barkana1999}; \citealp{SobacchiMesinger2013}), while \Rad\nspace captures a more gradual, outside-in reionization response that mitigates abrupt gas loss and suppression of accretion. HALO7 is a notable exception: despite being the most massive halo at $z=5$, its $M_{\rm gas}$ declines after $z \sim 7$, likely due to the combined effect of sustained Pop II-indiv CCSNe feedback at $z \gtrsim 7$ and the onset of reionization at $z=7$; we return to this case in Section \ref{EVOL}.

\par
However, a larger retained gas reservoir does not necessarily translate into a larger stellar mass. Among the six halos with higher $M_{\rm gas}$ in \Rad, HALO4 and HALO6 show stellar masses comparable to their \BG\nspace counterparts. In HALO4, much of the additional gas does not meet our star formation criteria ($n_{\rm H} \ge 10^{2} \rm\,{cm}^{-3}$ and $T_{\rm gas}<10^{4}\,\rm{K}$). This is mainly due to the interplay between Pop II-indiv SN feedback and reionization around $z\sim7$, as described in Section \ref{EVOL}, which heats and disperses the gas and prevents it from satisfying the star formation criteria. In HALO6, star formation continues in \Rad, but it is dominated by the massive Pop II-indiv stars (CCSN progenitors) which, by our definition (we track $M_{\star}$ only from Pop II-SSP, See Section \ref{Sec:starformationPopII}), do not contribute to $M_{\star}$. In contrast, the remaining four halos form substantially more stars in \Rad, reaching final stellar masses that are typically an order of magnitude higher than in \BG.

\subsubsection{Star-forming gas \& Star formation history}

\begin{figure*}
    \centering
    \includegraphics[width = 175mm]{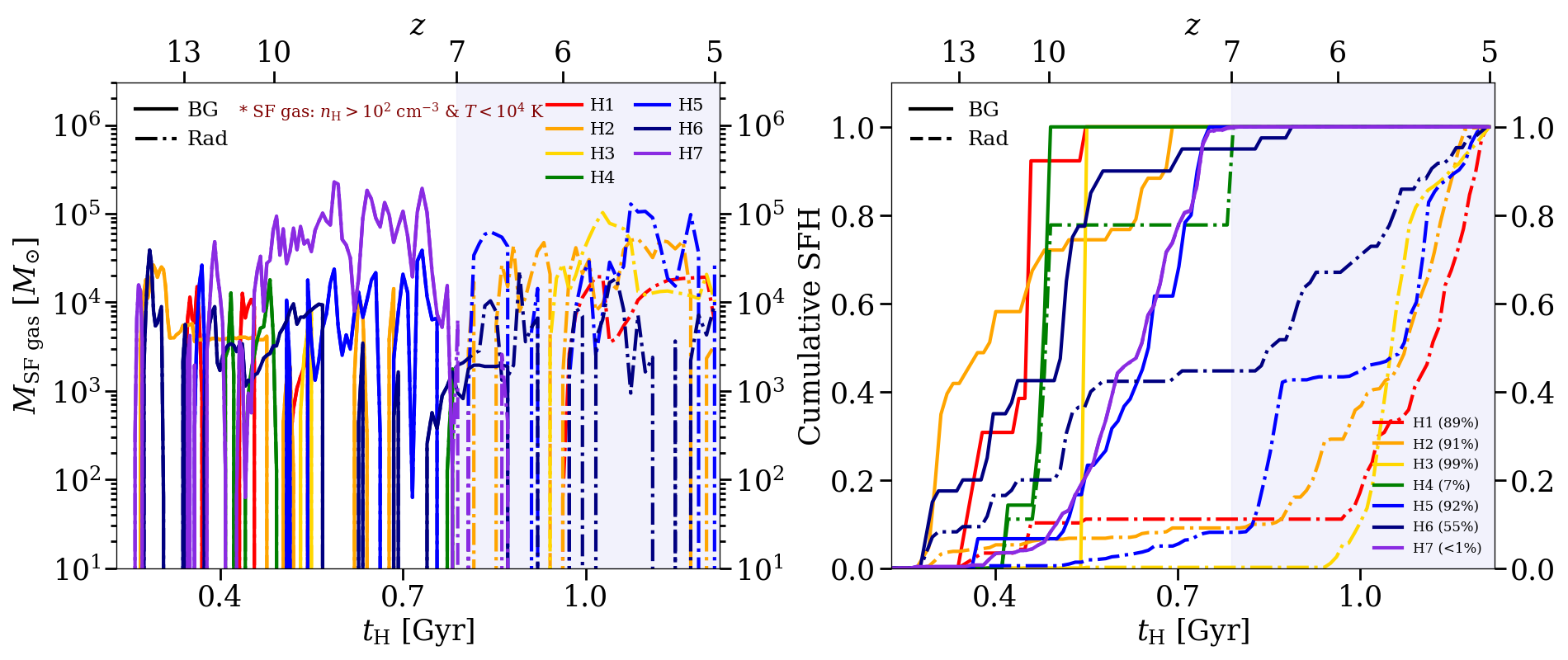}
    \caption{\texttt{Left}: Mass evolution of star-forming gas, which is defined by the criteria of a neutral hydrogen number density  $n_{\rm{H}}>10^{2} \, \rm{cm}^{-3}$ and a temperature $T<10^{4} \, {\rm{K}}$. In the \BG\nspace runs, all the halos except HALO6 lose their star-forming gas after the onset of reionization ($z\le7$). In contrast, the \Rad\nspace runs retain more than $10^{3}\Msun$ of star-forming gas even at $z=5$, except HALO4 and HALO7. \texttt{Right}: Cumulative SFHs (cSFHs) of each halo. To compare the intrinsic star-forming capability of each system, we only include in-situ star formation. In all the \BG\nspace runs, at least 90\% of the stars form before the onset of reionization, indicating that the homoegeneous reionization background effectively suppresses subsequent star formation. By contrast, the \Rad\nspace runs, with the exception of HALO4 and HALO7, exhibit extended star formation down to $z=5$.}
    \label{fig:sfgassfh}
\end{figure*}

To connect the global mass growth trends described in the previous section to the actual star formation activity in each halo, we next examine the evolution of the star-forming gas reservoir and resulting SFHs. Figure \ref{fig:sfgassfh} links the gas conditions for star formation to the resulting SFHs in the target UFDs. The left panel tracks the mass of the star-forming gas, defined by the same criteria used for star formation in the simulation ($n_{\rm{H}}>10^{2} \, \rm{cm}^{-3}$ and $T<10^{4} \, {\rm{K}}$). The right panel shows cumulative star formation histories (cSFHs) for in-situ stars only, so that the curves reflect the intrinsic star formation activity of each halo rather than ex-situ contributions. We caution that the apparent convergence of some HALO-\Rad s \nspace cSFHs to 1.0 at $z=5$ reflects the termination of our simulations, rather than genuine quenching.

\par
As indicated in the left panel of Figure \ref{fig:sfgassfh}, the prescription \BG \nspace rapidly removes the star-forming gas reservoir after the onset of reionization. In \BG, all halos except HALO6 lose essentially all star-forming gas shortly after $z=7$, and the subsequent evolution remains close to zero. In \Rad, on the other hand, the reservoir of star formation gas persists to later times: most halos retain $\gtrsim 10^{3}\Msun$ of star formation gas at $z=5$, indicating that dense, cool gas survives within the halo despite the progress of ionizing radiation. HALO4 and HALO7 are exceptions in \Rad, where the star-forming gas remains strongly depleted even toward $z=5$. This is because HALO7 is fully quenched before the onset of reionization by cumulative SN feedback, while in HALO4 a brief reignition episode at $z \approx 7.0$ fails to rebuild a lasting star-forming reservoir (Section \ref{EVOL}).

\par
The cSFHs in the right panel of Figure \ref{fig:sfgassfh} show how these differences in star-forming gas lead to distinct star formation outcomes. In \BG, the cSFHs rise steeply before $z=7$ and then become nearly flat, showing that star formation is effectively truncated at the onset of reionization in every halo. In \Rad, the cSFHs exhibit multiple shapes rather than a single synchronized pattern: six of the seven halos show extended star formation relative to their \BG\nspace counterparts ($\Delta \tau_{90} =300-740 \, \rm{Myr} $; Table \ref{tab:tau90}). Five of these (HALO1, HALO2, HALO3, HALO5, and HALO6) continue forming stars down to $z=5$, while in HALO4 the extension is limited to a brief reignition episode shortly after the onset of reionization ($z \approx 7.0$). Only HALO7, fully quenched before the onset, shows no delay. This diversity is quantified by the labeled fractions for the \Rad\nspace runs, which give the fraction of stars formed at $z<7$ (e.g., 89\% in HALO1, 91\% in HALO2, 99\% in HALO3, 7\% in HALO4, 92\% in HALO5, 55\% in HALO6, and $<1\%$ in HALO7). This halo-to-halo diversity arises even though the target halos span a narrow mass range at $z=5$ ($M_{\rm vir} \sim 10^{7-8} \Msun$; shown in Table \ref{tab:set}), indicating that the EoR environment---captured only by realistic, patchy reionization modeling such as our \Rad\nspace implementation---can be as important as halo mass in shaping stellar mass assembly of UFDs. 

\begin{table}
\centering
{\setlength{\tabcolsep}{12pt}
\begin{tabular}{c | c | c | c}
\hline
\hline
Name & ${\Delta}t_{\tau_{90, \,\rm{{\nspace}Rad-BG}}}$ [Myr] & $z_{\tau_{90, \,\rm{BG}}}$ & $z_{\tau_{90, \,\rm{Rad}}}$\cr
\hline
{\sc HALO1} & 740 & 10.47 & 5.04 \cr
{\sc HALO2} & 480 & 7.82 & 5.17 \cr
{\sc HALO3} & 600 & 9.16 & 5.21 \cr
{\sc HALO4} & 300 & 9.97 & 6.98 \cr
{\sc HALO5} & 430 & 7.40 & 5.17 \cr
{\sc HALO6} & 540 & 8.85 & 5.33 \cr
{\sc HALO7} & 0 & 7.33 & 7.33 \cr
\hline
\hline
\end{tabular}}
\caption{Information about the time when 90\% of the star formation has been completed (${\tau_{90}}$). Column (1): Name of run. Column (2): Difference in $\tau_{90}$ between \BG\nspace and \Rad. Column (3): Redshift corresponding to $\tau_{90}$ in \BG\nspace runs. Column (4): Redshift corresponding to $\tau_{90}$ in \Rad\nspace runs. Because several \Rad\nspace halos are still forming stars at the end of the simulation ($z=5$), the quoted $\tau_{90}$ values in \Rad\nspace may be reached later in a longer run. Accordingly, the inferred $\Delta \tau_{90}$ values may represent lower limits.}
\label{tab:tau90}
\end{table}

\par
Table \ref{tab:tau90} summarizes the redshift at which $90\%$ of the cumulative star formation is completed ($\tau_{90}$) and the corresponding delay between \Rad\nspace and \BG, $\Delta{t_{\tau_{90}, \,\rm{Rad - BG}}}$ (we also report $\tau_{80}$ and $\tau_{50}$ in Section \ref{qtimes} for observational comparison). Averaged over the full sample, star formation in \Rad\nspace is extended by $\langle \Delta {t_{\tau_{90}, \,\rm{Rad - BG}}} \rangle \approx 440 \,\rm{Myr}$ relative to \BG. We note again that because several \Rad\nspace halos are still forming stars at the end of the simulations ($z=5$), the inferred $\tau_{90}$ values in \Rad\nspace may be underestimated. The averaged delay time quoted here may, therefore, also represent a lower limit. Taken together with Figure \ref{fig:sfgassfh}, these trends support a consistent trend within our sample: the \Rad\nspace scheme retains more star-forming gas inside halos through self-shielding, enabling extended and system-dependent SFHs, whereas the \BG\nspace runs more readily lose the star-forming reservoir and exhibit more uniform early truncation.

\par
Consistent with the delayed and diverse SFHs in the \Rad\nspace runs, observed LG UFDs do not show a single, universal truncation at a fixed redshift, but instead exhibit system-to-system diversity in their quenching times. Our \Rad\nspace runs produce a mean quenching delay of $\sim 440\,\rm{Myr}$ relative to the \BG\nspace runs. This delay is comparable to the delays inferred for observed UFDs, including $\sim 600\,\rm{Myr}$ for LMC satellites relative to other MW satellites in \cite{Sacchi_2021}, and up to $\sim 800,\rm{Myr}$ for LMC satellites and first-infall systems relative to long-term MW satellites in \cite{Meredith2025} based on $\tau_{80}$. Thus, our work suggests that differences in the local EoR environment, such as variations in the timing and intensity of patchy reionization, may be one possible origin of the observed system-to-system scatter in UFD quenching times.

\subsubsection{Radial distribution of star formation}
\label{locationofsf}

\begin{figure*}
    \centering
    \includegraphics[width = 135mm]{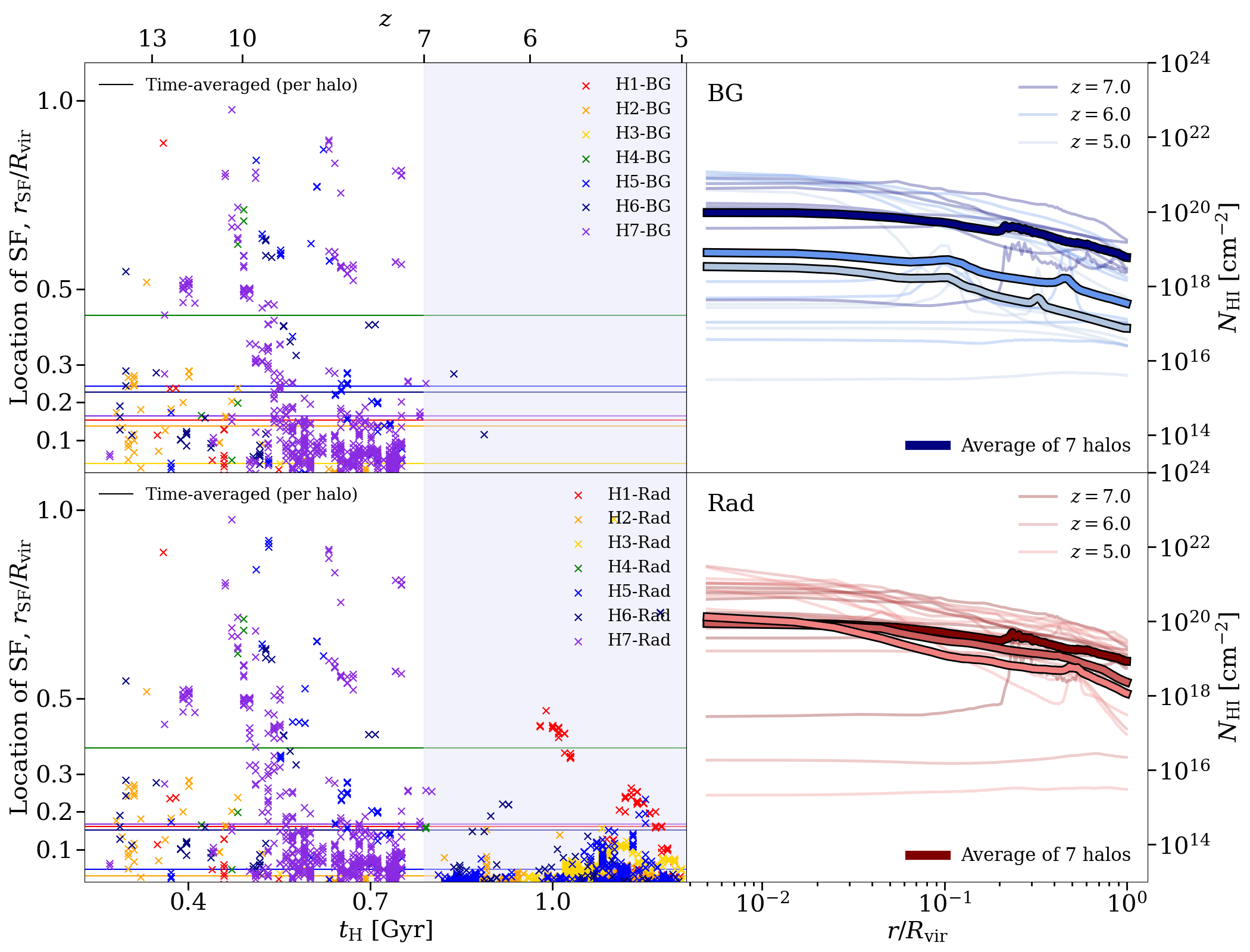}
    \caption{\texttt{Left panels}: The location of the star formation events in each run. In the \Rad\nspace runs (lower panel), before redshift 7, star-forming regions are distributed throughout the inner $R_{\rm{vir}}$, whereas after redshift 7, they are concentrated within $0.3R_{\rm{vir}}$. This indicates that due to self-shielding, only the gas in the central region ($r<0.3R_{\rm{vir}}$) remains cold and dense enough to meet the star formation criteria. All HALO-{\BG}s do not form any stars after redshift 7, except HALO6. \texttt{Right panels}: Column density of HI gas as a function of distance from the halo center. All seven halos are overplotted without distinction, and different colors indicate different redshifts. Thick solid lines show the average of the seven halos at each redshift. In the \BG\nspace case, the column density profiles of most halos show only a weak central rise by $z=5$, whereas in the \Rad\nspace case the column density increases toward the center for most halos, reaching peak values that are about 1.5 orders of magnitude higher than in \BG. This indicates that, only in the \Rad\nspace runs, the dense central gas can survive until $z=5$ due to self-shielding.}
    \label{fig:sfregion}
\end{figure*}

Building on the evolution of the star-forming gas reservoir and the resulting SFHs, we next examine where star formation actually takes place within the halos and how the neutral gas is distributed with radius. This analysis helps us understand why extended star formation emerges preferentially in the \Rad\nspace runs. In Figure \ref{fig:sfregion}, the left panels show the birth radii of in-situ star formation events, $r_{\rm SF}$, normalized by the virial radius of the host halo, $R_{\rm vir}$, as a function of cosmic time. Individual symbols mark the locations of star formation events, while the horizontal solid lines indicate the time-averaged value of $r_{\rm SF}/R_{\rm vir}$ for each halo.

\par
In the \Rad\nspace runs (lower left), star formation becomes extremely centrally concentrated after the onset of reionization: at $z < 7$, $ \sim99\,\%$ of in-situ star formation events occur within $0.3 R_{\rm vir}$ (only $\sim1\,\%$ occur at larger radii) across the seven halos. This implies that star-forming gas is confined to the inner halo. There, reionization proceeds in an outside-in manner, so the dense central gas can remain cold and neutral due to self-shielding, allowing star formation to continue. In the \BG\nspace runs, however, the UVB is imposed on the entire halo simultaneously, so the central reservoir is not shielded by an outer neutral layer and its density gradually falls below the star formation threshold. As a result, star formation shuts down after $z=7$ in all \BG\nspace halos except HALO6, which exhibits only two star formation events between $7 \lesssim z \lesssim 6$.

\par
The right panels of Figure~\ref{fig:sfregion} further support this interpretation by showing the density of the column $\rm{H\,I}$ $N_{\rm HI}$, as a function of radius. Here, $N_{\rm HI}$ is computed by integrating the neutral hydrogen number density along the line of sight ($z$-axis), providing a direct measure of the self-shielding capability of gas at a given projected radius. We overplot the seven halos without distinguishing them by color; instead, color indicates redshift. We show the mean $\rm{H\,I}$ column density across the seven halos as thick solid lines at each redshift. In the \BG\nspace case (upper right), the central rise of $N_{\rm HI}$ is much weaker by $z=5$, showing that the neutral column in the halo center is substantially reduced once the background is established. In the \Rad\nspace case (lower right), $N_{\rm HI}$ generally rises toward the center and reaches peak values that are $\sim 1.5$ dex higher than in \BG\nspace at $z=5$, demonstrating that a dense, neutral central reservoir can persist to late times. Therefore, the strong central concentration of star formation and the survival of a high $N_{\rm HI}$ core naturally explain why extended SFH features appear preferentially in the \Rad\nspace runs.

\subsection{Case studies of evolutionary pathways}
\label{EVOL}

\begin{figure*}
    \centering
    \includegraphics[width = 175mm]{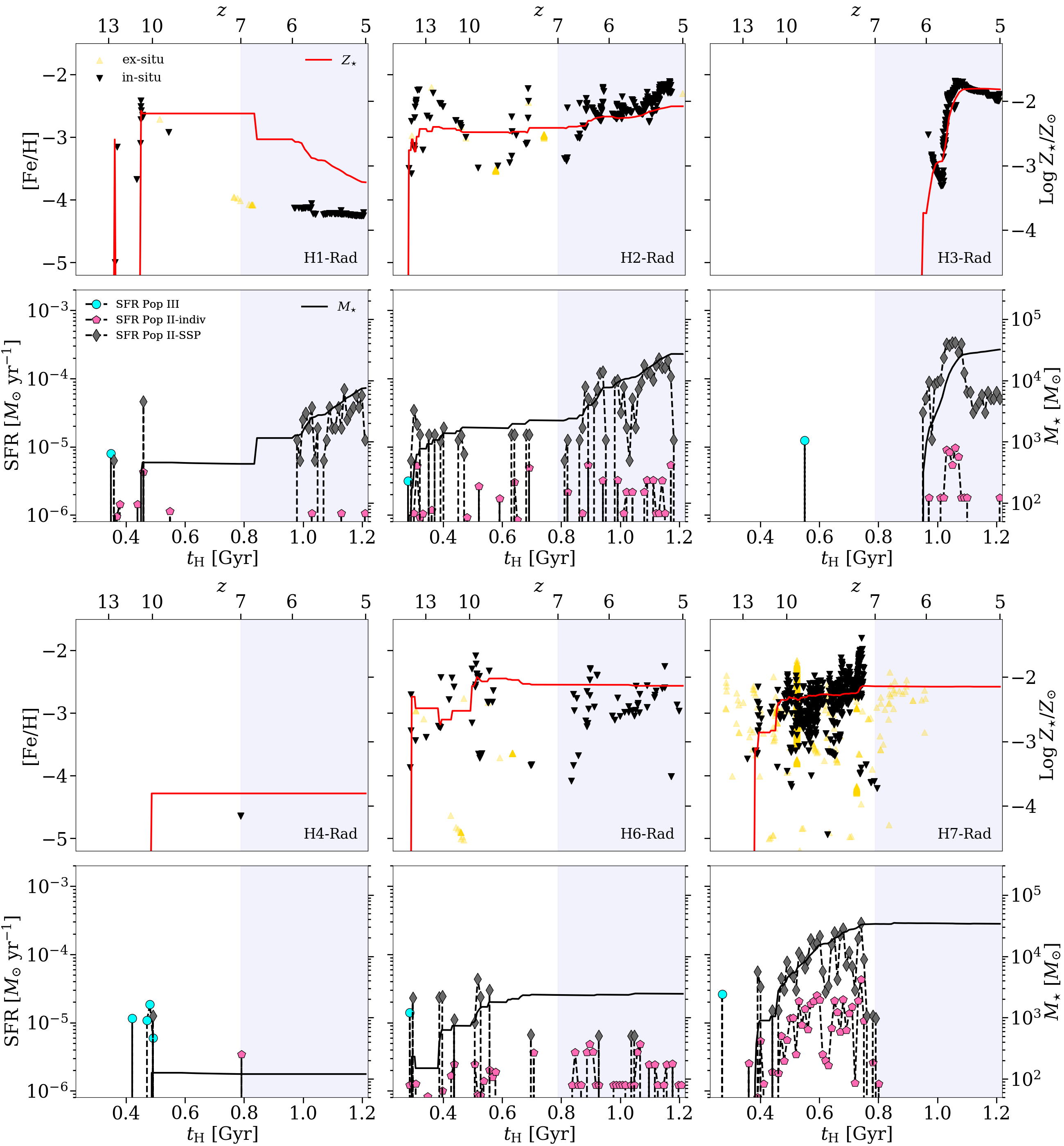}
    \caption{Stellar evolution in each \Rad\nspace run. For each halo, we show two panels: the upper panel presents metallicity, and the lower panel shows star formation rate (SFR) and stellar mass evolution. In the upper panels, the red solid line indicates the mean stellar metallicity, while the (inverted) triangles show the iron-to-hydrogen ratio of individual star particles. In-situ stars are shown as black inverted triangles, while ex-situ stars are shown as yellow triangles. The lower panels show the evolution of stellar mass (black solid line) and the in-situ SFR (black dashed line). Cyan circle, pink pentagon, and gray diamonds indicate the SFR of Pop III, Pop II-indiv, and Pop II-SSP, respectively. We omit HALO5 because its evolutionary trend is very similar to that of HALO2. The halos shown in this figure exhibit diverse evolutionary histories. H1-\Rad: during a temporary quenching phase ($\sim400$Myr), metals are efficiently diluted on inter-stellar medium (ISM) scales, so newly formed stars have [Fe/H] values that are $\sim 1$ dex lower than the previous generation. H2-\Rad: sustained star formation produces a monotonic increase in both stellar mass and mean stellar metallicity. H3-\Rad: two distinct star-forming clouds cause a decrease in the mean stellar metallicity after $z\approx5.8$, even though star formation appears continuous. H4-\Rad: intense Pop III feedback suppresses subsequent star formation. H6-\Rad: star formation continues down to $z=5$, but most Pop II stars formed after the onset of reionization are massive and short-live (Pop II-indiv), so they do not contribute to the surviving stellar mass. H7-\Rad: SN feedback from numerous Pop II stars eventually shuts down further star formation.}
    \label{fig:stellarevolution}
\end{figure*}

As shown in the previous sections, the \Rad\nspace reionization model produces extended star formation in six UFDs (five of which sustain it to $z=5$). In this section, we examine whether these late-forming stars---particularly those formed after the onset of reionization at $z=7$---exhibit stellar properties distinct from those of earlier populations. In addition, we investigate why HALO4 fails to sustain its brief post-reionization episode and HALO7 quenches before the onset. Figure \ref{fig:stellarevolution} consists of two panels for each halo. The upper panel shows the mean stellar metallicity (red solid line) and the iron-to-hydrogen ratios of individual star particles, with in-situ (black inverted triangle) and ex-situ (yellow triangle) stars distinguished by symbol shape and color. The lower panel presents the in-situ star formation rate (SFR; black dashed line) and the stellar mass (black solid line). Different colors and symbols in SFR indicate different stellar particle types: Pop III (cyan circles), Pop II-indiv (pink pentagons), and Pop II-SSP (gray diamonds). We omit HALO5 because its evolutionary pattern is similar to that of HALO2.

\par
In fact, our simulated HALO-\Rad s exhibit diverse evolutionary histories. We therefore discuss each halo individually rather than grouping them by a single property. Before presenting the detailed analysis, we note that all HALO-\BG s except HALO6 quench rapidly after the onset of reionization ($z=7$); consequently, no data points appear at $z<7$ in either panel because no stars form. We therefore show only the HALO-\Rad\nspace results. We also note that most of the chemical enrichment in our simulations is driven by CCSNe from Pop III and Pop II-indiv stars, with relatively small contributions from Type Ia SN and AGB stars (from Pop II-SSP stars). Separately, we compute the mean stellar metallicity and the stellar mass evolution using only Pop II-SSP stars. For details of our feedback implementation, see Section \ref{STELLARFEEDBACK}.

\par
HALO1-\Rad: From $z \approx 9.2$ to $z \approx 6.0$, in-situ star formation is temporarily quenched ($\sim 400\, \rm{Myr}$). During this quenching phase, the system produces no new metals through ongoing star formation, while previously enriched metals from early stars ($9 \lesssim z \lesssim 13$) have sufficient time to mix on ISM scales. At the same time, the gas-phase metallicity is reduced by dilution as metal-poor gas is (re)accreted and mixed into the ISM. When in-situ star formation resumes at $z \lesssim 6$, newly formed stars therefore form from this diluted, metal-poor gas and reach $[\rm{Fe/H}] \sim -4$, about $\sim 1$ dex lower than the early stellar population. As a result, the mean stellar metallicity decreases by $\sim 1$ dex during the EoR ($5 \lesssim z \lesssim 7$), even though HALO1-\Rad\nspace continues forming stars and its stellar mass grows by $\sim$ an order of magnitude relative to its \BG\nspace counterpart. This case highlights that galaxy growth does not necessarily imply a monotonic increase in stellar metallicity in UFD regimes: a temporal quenching episode can decouple mass growth from chemical enrichment through dilution.

\par
HALO2-\Rad\nspace (HALO5-\Rad): After the first star forms, star formation proceeds continuously, leading to monotonic increases in both stellar mass and mean stellar metallicity. We attribute this sustained star formation to two factors: (i) a larger halo mass at the epoch of first star formation ($z \sim 13$), about three times that of HALO1-\Rad, and (ii) a lower Pop III SFR, also $\sim 3$ times lower than in HALO1-\Rad. These conditions make the system more resilient to early stellar feedback, allowing star formation to continue without a prolonged interruption. 

HALO3-\Rad: Following strong Pop III feedback, star formation is temporarily quenched for $\sim 400\, \rm{Myr}$. After this period, the star-forming gas recovers, and star formation then reignites at $z \lesssim 6$ and continues thereafter, so the stellar mass increases monotonically. The individual stellar $[\rm {Fe/H}]$ values rapidly increase from $[\rm {Fe/H}] \sim -4$ to $\sim -2$ after star formation resumes, but then show a flatter evolution, or a slight decrease, around $z \sim 5.6$. From the gas morphology and gas-phase metallicity distribution, we find that this behavior reflects star formation in two gas clouds with different metallicities within the UFD.

\par
HALO4-\Rad: In most of our UFDs, Pop III star formation occurs in a single brief episode, with a peak SFR of $\sim 10^{-5} \Msun \, \rm{yr}^{-1}$. In contrast, HALO4-\Rad\nspace shows four such Pop III SFR peaks, implying $\sim 4 \, \times$ more Pop III star formation episodes than the other systems. The resulting strong Pop III feedback efficiently heats and expels the dense, cold gas, leaving no star-forming reservoir immediately afterward. Star formation can resume only after the system re-accretes cool gas, which occurs after $\sim 300 \, \rm{Myr}$. However, this recovered gas forms Pop II-indiv stars; their subsequent CCSNe again evacuate the remaining star-forming gas. The halo therefore fails to re-enter a sustained star-forming phase and becomes quenched. Because star formation truncates early, HALO4-\Rad\nspace provides limited leverage to track long-term evolution in stellar mass and metallicity. Together with HALO7-\Rad\nspace, it serves as an example of quenching driven primarily by internal feedback in UFDs.

\par
HALO6-\Rad: Before the onset of reionization, HALO6-\Rad\nspace follows a monotonic increase in stellar mass and mean stellar metallicity, similar to HALO2-\Rad\nspace (HALO5-\Rad). After $z=7$, these quantities remain nearly constant while star formation continues. This behavior arises because late-time star formation is dominated by Pop II-indiv stars, which are therefore excluded from our calculations of stellar mass and mean stellar metallicity. Despite adopting a Salpeter IMF over $0.1-100\,\Msun$, we find that the Pop II-indiv mass range ($8-40\,\Msun$) is over-represented relative to the Pop II-SSP component ($0.1-8\,\Msun$) compared to the analytic IMF expectation. This suggests a mismatch between the intended IMF sampling and the effective stellar population realized in our simulations, leading HALO6-\Rad\nspace to exhibit an evolutionary pattern distinct from the other halos. While these diagnostics suggest little or no evolution in stellar mass and mean stellar metallicity, we emphasize that star formation persists after $z=7$, indicating an extended star formation episode.

\par
HALO7-\Rad: As the most massive halo in our sample---approximated $4-10$ times more massive than the others---HALO7 undergoes a higher frequency of early star formation. However, the excessive formation of Pop II-indiv stars drives a large number of CCSNe, rapidly heating the halo gas and expelling the star-forming gas reservoir. Once the gas is evacuated, cosmic reionization after $z=7$ further suppresses cooling and gas re-accretion, preventing the system from re-entering a star-forming phase. In other words, its evolution is characterized by an early starburst (at $z \gtrsim 7$) and a prolonged quenching phase until the end of reionization. Together with the HALO4-\Rad\nspace result, this suggests that star formation in low-mass UFDs can be quenched not only by cosmic reionization but also by internal SN feedback, with reionization subsequently inhibiting any late-time recovery.

\par
In the \Rad\nspace runs, six of the seven halos show delayed quenching relative to their \BG\nspace counterparts, and five of them sustain star formation down to $z=5$. HALO7 is quenched before the onset of reionization and shows no delay, while in HALO4 the delay reflects only a brief reignition episode at $z\approx7.0$. Notably, post-reionization ($z \lesssim 7$) star formation activities of the five sustained systems diverge substantially, with no single, unified evolutionary trend. These results emphasize that galactic properties of UFDs reflect a complex interplay between the external radiation field and internal physical processes, including stellar feedback, gas-phase conditions, and the virial mass at the epoch of feedback. We therefore caution that interpreting the observed diversity of UFD properties requires considering a range of stochastic evolutionary pathways.

\subsection{Observable quantities}
\label{OBSERVABLE}
\subsubsection{Stellar mass-metallicity relation}

\begin{figure}
    \centering
    \includegraphics[width=1.0\linewidth]{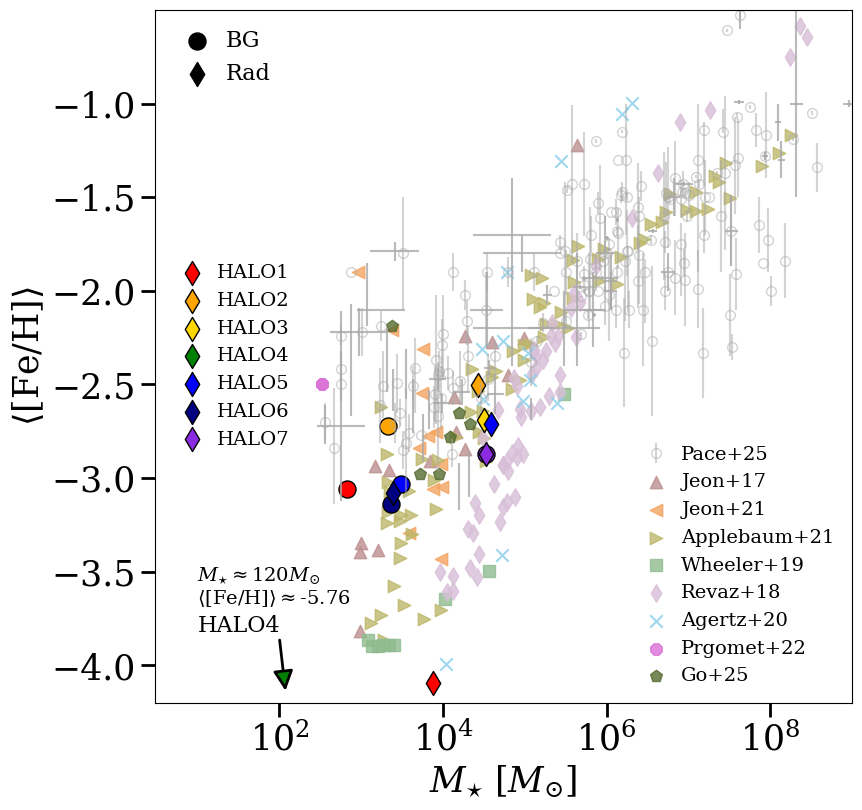}
    \caption{Stellar mass versus averaged iron-to-hydrogen ratio (MZR) of each halo, shown in different colors with circles (\BG) and diamonds (\Rad). The faint-colored symbols represent results from simulations, while the gray open circles indicate those from observations. Among the seven halos, five HALO-\Rad s \nspace exhibit extended star formation down to $z=5$, but only three of them (HALO2, HALO3, and HALO5) exhibit higher stellar masses and higher mean stellar metallicities. This indicates that there is no single clear trend separating \BG\nspace and \Rad\nspace in the MZR. Nevertheless, our results remain in good agreement with previous simulations of UFDs and are also consistent with observed UFDs at higher stellar masses ($M_{\star}>10^{4}\Msun$).}
    \label{fig:MZR}
\end{figure}

\begin{figure*}
    \centering
    \includegraphics[width = 175mm]{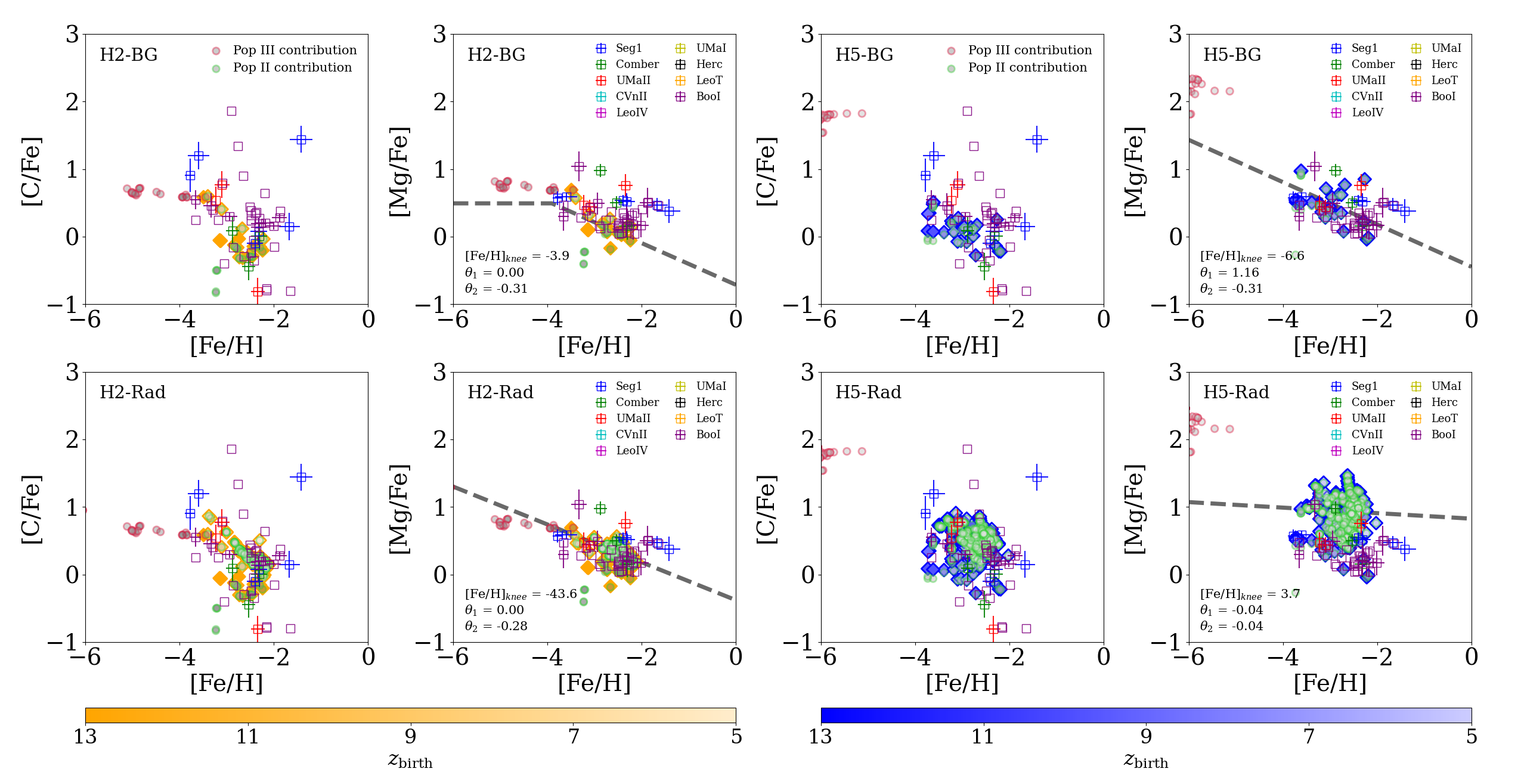}
    \caption{Stellar abundances of two simulated galaxies at $z=5$. Stars from each galaxy are plotted with a consistent base color (HALO2 in orange and HALO5 in blue), with different colors indicating different formation times. Gray circles with red/green edges indicate Pop III/Pop II enrichment contributions, respectively. For each galaxy, the upper panels show the \BG\nspace run and lower panels show the \Rad\nspace run; the left panels display [C/Fe] vs. [Fe/H], and the right panels display [Mg/Fe] vs. [Fe/H]. Our results reproduce the stellar abundance patterns of observed UFDs reasonably well, but do not exhibit a clear abundance trend with stellar age; instead, the stars are broadly scattered over $-4<[\rm{Fe/H}]<-2$. Although star formation is clearly more extended in the \Rad\nspace runs, this is insufficient to imprint a pronounced metallicity difference, such as a distinct $\alpha$-element ``knee".}
    \label{chemabundances}
\end{figure*}

In general, stellar mass and stellar metallicity are positively correlated: more massive galaxies typically form more stars, produce more metals, and retain them more efficiently. In the UFD regime, however, the mass--metallicity relation (MZR) exhibits substantial scatter, and the correlation becomes weak at the lowest stellar masses (e.g., \citealp{Kirby2013, Simon2019, Go2025, Pace2025}). Moreover, at fixed $M_{\star}$, theoretical predictions for the mean stellar metallicity, $\langle \rm{[Fe/H]} \rangle$, are often $0.5-2$ dex lower than observational estimates in this regime (e.g., \citealp{Wheeler2019, Agertz2020, Sanati2023, Go2025}).

Many simulation studies have attempted to explain the origin of this discrepancy (e.g., \citealp{Jeon_2017}; \citealp{Wheeler2019}; \citealp{Agertz2020}; \citealp{Applebaum2021}; \citealp{Prgomet2022}; \citealp{Sanati2023}; \citealp{Go2025}), but it remains under debate. We therefore examine the MZR of our simulated halos, focusing on how well our results reproduce the observed trend and how they compare with previous simulations. For observational comparison, we adopt the recent compilation of nearby dwarf galaxies (including UFDs) and star clusters assembled in the Local Volume Database (LVDB; \citealp{Pace2025}).

\par
Figure \ref{fig:MZR} shows the MZR for each halo, with results from other studies. Here, $M_{\star}$ and $\langle {\rm [Fe/H]} \rangle$ are computed using all stars contained in each system at $z=5$, including both in-situ and ex-situ populations. We first examine the trends exhibited  by our simulated halos, showing the $z=5$ results as circles for \BG\nspace and diamonds for \Rad. Compared with \BG, \Rad\nspace shows a tendency toward higher stellar masses and higher stellar metallicities, particularly in HALO2, HALO3, and HALO5. The remaining halos show different changes, for reasons related to their individual evolutionary paths (see Section \ref{EVOL}); for example, some systems are quenched by internal feedback (HALO4 and HALO7), while dilution during temporary quenching lowers $\langle{\rm [Fe/H]}\rangle$ (HALO1).

\par
Next, we compare our halos with other works. Our results fall within the range of previous simulation predictions (e.g., \citealp{Jeon_2017}; \citealp{Applebaum2021}; \citealp{Jeon2021b}; \citealp{Go2025}). At relatively large stellar masses ($M_{\star} \gtrsim 10^{4} \Msun$), our halos are also broadly consistent with the observations. At lower stellar masses ($M_{\star} \lesssim 10^{4} \Msun$), however, we reproduce the same qualitative tension reported in the literature: the observed UFDs include systems with $\langle \rm{[Fe/H]} \rangle \gtrsim -2.5$, while our halos tend to extend to lower mean metallicities, roughly $-5 \lesssim \langle \rm{[Fe/H]} \rangle \lesssim -2.5$.

\subsubsection{Chemical abundances}
SNe inject both energy and metal yields into the ISM, fundamentally shaping the chemical composition of subsequent stellar generations. Expanding on our analysis of iron abundance, we now examine the relative ratios of carbon and magnesium compared to iron. Carbon serves as a sensitive diagnostic of early enrichment, which can highlight contributions of Pop III, while magnesium provides a robust $\alpha$-element tracer of CCSN activity. Magnesium is particularly informative because $\alpha$-element abundances encode star formation timescales; prolonged star formation allows Type Ia SNe to contribute substantial iron, thus lowering [$\alpha$/Fe] and producing a characteristic ``knee" in the [$\alpha$/Fe]-[Fe/H] plane (e.g., \citealp{Tolstoy2009}). Given that our \Rad\nspace runs exhibit a mean delayed quenching time of $\sim 440\, \rm{Myr}$ relative to \BG, we investigate whether this extended duration is sufficient for such a knee-like feature to emerge.

\par
Figure \ref{chemabundances} presents the [C/Fe]-[Fe/H] and [Mg/Fe]-[Fe/H] ratios of HALO2 and HALO5 using stars within $R_{\rm vir}$ at $z=5$; colors indicate stellar formation times. We choose these two systems because they continue to build up stellar mass and increase their mean stellar metallicity (Section \ref{EVOL}), enabling a more informative abundance analysis. To separate enrichment channels, we additionally show gray circles with red/green edges to mark Pop III/Pop II enrichment contributions. For a direct comparison with observations, we do not distinguish in-situ and ex-situ stars. Observed LG UFD abundances are overplotted as open squares with 1$\sigma$ uncertainties, compiled from \cite{Vargas2013} for Seg 1, Leo IV, Leo T, Comber, UMa I, UMa II, CVn I, CVn II, and Herc; \cite{Gilmore2013} and \cite{Frebel2016} for Boo I; \cite{Frebel2014} for Seg 1; and \cite{Frebel2010} for UMa II and Comber.

\par
Overall, our simulated abundance patterns are broadly consistent with those of observed LG UFDs: both [C/Fe] and [Mg/Fe] for HALO2 and HALO5 fall within the range spanned by the observational compilations. Motivated by the expectation that early Pop III CCSNe (around $z \sim 13$) could imprint carbon-enhanced signatures, we anticipated a higher fraction of carbon-enhanced metal-poor (CEMP) stars---commonly defined as [C/Fe] $ \ge 0.7$ and [Fe/H] $ \le -2$ (e.g., \citealp{Aoki2007})---among the earliest formed populations. Separating the enrichment channels shows that Pop III contributions can indeed produce CEMP-like abundances, but we find no stars at $z=5$ whose chemistry reflects a purely Pop III signatures. Instead, subsequent Pop II enrichment washes out any obvious correlation between stellar birth time and [C/Fe]. Interestingly, in the \Rad\nspace runs of both halos, stars formed after the onset of reionization populate a wide metallicity range ([Fe/H] $\sim -4$ to $-2$) yet typically exhibit [C/Fe] values higher by $\sim 0.5$ dex than the pre-reionization populations.

\par
The systematically higher [C/Fe] among stars formed after $z \sim 7$ is naturally explained by the fact that star formation at these epochs is dominated by Pop II stars. As Pop II star formation proceeds, delayed enrichment from the Pop II-SSP stars---representing intermediate-mass AGB progenitors in our model---becomes increasingly important, injecting substantial carbon while contributing little iron. This interpretation is supported by our channel-separated enrichment estimates: stars with [C/Fe] $> 0.5$ also appear when considering Pop II contributions alone, indicating that Pop II enrichment via AGB ejecta can generate CEMP-like abundances without requiring a purely Pop III origin.

\par
We expected that the extended star formation in the \Rad\nspace runs might allow an increasing contribution from Type Ia SNe even in UFD-scale systems, potentially producing a "knee" in the [Mg-Fe]-[Fe/H] plane. However, we find no clear evidence for this feature, either visually or from quantitative fitting. We fitted a broken linear relation to [Mg/Fe] as a function of [Fe/H] using a piecewise model and non-linear least squares (e.g., \citealp{Ramirez2012}). However, the inferred knee parameters are not statistically meaningful, indicating that Type Ia enrichment remains sub-dominant despite the more extended star formation in \Rad. Nevertheless, our simulated [Mg/Fe] broadly agrees with the observed abundance patterns of LG UFDs.

\par
An exception is found in HALO5-\Rad, where stars formed after $z \le 7$ span a wide metallicity range ($-4 \lesssim $ [Fe/H] $\lesssim -2$) yet show higher [Mg/Fe] by $\sim 0.5$ dex compared to the pre-reionization ($z > 7$) population. This offset arises from difference in the Pop II CCSN progenitor mass distribution. In HALO2-\Rad, all post-reionization progenitors have $m \lesssim 30 \, \Msun$ and cluster near $10\, \Msun$, whereas in HALO5-\Rad, 42.3\% of progenitors exceed $30\,\Msun$. For the \cite{Portinari1998} yields at the relevant metallicities, $m \gtrsim 30\, \Msun$ progenitors eject less iron (by $\sim 2\, \times $) and more magnesium (by $\sim 4\,\times $) than $\sim 10\,\Msun$ progenitors, which increases [Mg/Fe]. Although such cases are rare, this indicates that extended star formation in a UFD can yield enhanced $\alpha$-element ratios if a large fraction of massive stars form at late times.

\par
In summary, among the halos that exhibit extended star formation, the two systems that continue to grow in both stellar mass and mean stellar metallicity reach carbon and magnesium abundance levels broadly comparable to those observed in LG UFDs. Although we do not identify a clear $\alpha$-element knee in the [Mg/Fe]-[Fe/H] plane, our results suggest that late-time star formation after $z \sim 7$ can be accompanied by carbon enhancement driven by Pop II AGB ejecta, and, in rare cases, by elevated [Mg/Fe] associated with a top-heavier Pop II CCSN progenitor population. These behaviors are absent in the \BG\nspace runs, where star formation is truncated more uniformly.

\subsubsection{Detectability as DLAs}
\begin{figure*}
    \centering
    \includegraphics[width = 155mm]{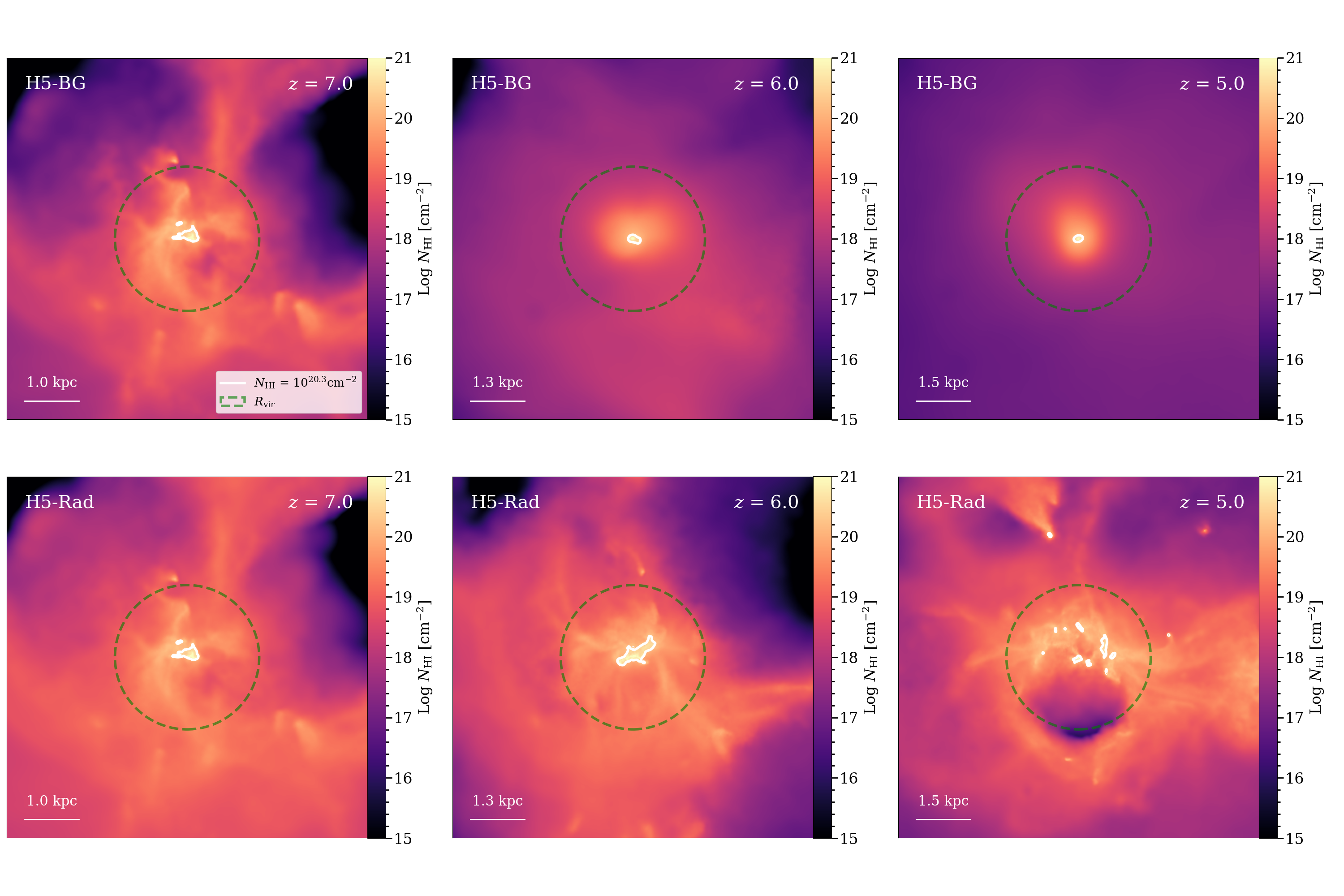}
    \caption{Projected HI column density maps. The top panels show the \BG\nspace run of HALO5, and the bottom panels show the corresponding \Rad\nspace run. We show HALO5 as an example; the other halos are summarized in the text. From left to right, the columns correspond to $z$=7, 6, and 5. The green dashed circle marks the virial radius of the halo, and the white contour at $N_{\rm{HI}}=10^{20.3}\,\rm{cm}^{-2}$ corresponds to the threshold for damped Lyman-$\alpha$ absorbers (DLAs) defined by \citet{Wolfe1986}. In the \BG\nspace run, only a single compact patch ($\sim 0.04 \, \rm{kpc}^{2}$ at $z=5$) survives above the DLA threshold after $z=7$, whereas in the \Rad\nspace run several patches with a total area of $\sim 0.13\, \rm{kpc}^{2}$ remain at $z=5$.}
    \label{fig:HIdensity}
\end{figure*}

Damped Lyman-$\alpha$ absorption systems (DLAs) are neutral gas absorption systems identified in quasar spectra, defined by $N_{\rm HI} \ge 10^{20.3} \, \rm{cm}^{-2}$ (\citealp{Wolfe1986}). Since DLA observations are based on absorption, they are not affected by galaxy light. DLAs therefore offer a simple way to assess whether a faint system can host a substantial neutral gas reservoir. In the context of UFD progenitors, producing a DLA-level column density requires a system to retain sufficiently neutral, dense gas against external photoionization and photoheating during reionization, which is expected to be difficult in shallow potential wells. Motivated by our finding that the \Rad\nspace prescription more effectively preserves star-forming gas than the \BG\nspace model, we examine whether \Rad\nspace can yield gas-rich, DLA-like configurations even in relatively low-mass halos. Here, we restrict our analysis to whether such gas-rich regions arise in our simulations, and refer readers to \citet{Jeon2019_DLA} for a focused discussion of UFD detectability as DLAs.

\par
Figure \ref{fig:HIdensity} shows projected maps of $N_{\rm HI}$ at $z=7$, $6$, and $5$, highlighting regions that exceed the DLA threshold. We present HALO5 as an example, with \BG\nspace in the upper row and \Rad\nspace in the lower row, and summarize the other halos below. The green dashed circle marks $R_{\rm vir}$ at each redshift, and the white contour denotes the DLA threshold, $N_{\rm HI}=10^{20.3} \rm{cm}^{-2}$. In HALO5-\BG, the DLA-level gas contracts to a single compact patch after $z=7$, consistent with photoheating by a spatially uniform UVB that drives gas expansion and lowers the density of the neutral gas. In HALO5-\Rad, by contrast, the DLA-level gas extends over a larger region at $z=6$ and survives as several patches at $z=5$. Summing the pixels above the threshold, the DLA-level area at $z=5$ is $\sim 0.13 \, \rm{kpc}^2$ in \Rad\nspace and $\sim 0.04 \, \rm{kpc}^{2}$ in \BG.

The same trend holds across the sample. At $z=5$, three \BG\nspace halos (HALO2, HALO5, and HALO6) retain a small DLA-level patch with areas of $0.005-0.08 \, \rm{kpc}^{2}$, while the other four contain none. In \Rad, six of the seven halos retain DLA-level gas, with areas of $0.01-0.4 \, \rm{kpc}^{2}$, and in the three halos where both schemes keep such gas, the \Rad\nspace area exceeds the \BG\nspace area by a factor of $4-40$. HALO7, quenched before the onset of reionization, contains no DLA-level gas in either scheme. Given such small physical cross-sections, the probability of intersecting structures is extremely low. The systematically larger area in \Rad, however, suggests that halos modestly more massive than those in our sample might sustain a larger DLA-level neutral gas area and, in turn, could become more readily observable as DLAs when modeled under the \Rad\nspace scheme.

\section{Discussion}
\label{Sec:Discussion}
\subsection{Interpreting UFD quenching times}
\label{qtimes}

\begin{table}
\centering
{\setlength{\tabcolsep}{7pt}
\begin{tabular}{c | c | c | c}
\hline
\hline
Quantity & $\langle \BG\nspace \rangle$ [Gyr] & $\langle \Rad\nspace \rangle$ [Gyr] & $\Delta t_{\tau,\,{\rm BG-Rad}}$ [Myr]\cr
\hline
{\sc $\tau_{90}$} & 13.15 & 12.70 & 441.4 \cr
{\sc $\tau_{80}$} & 13.16 & 12.74 & 421.8 \cr
{\sc $\tau_{50}$} & 13.22 & 12.85 & 374.5 \cr

\hline
\hline
\end{tabular}}
\caption{Comparison of quenching times between the \BG\nspace and \Rad\nspace runs. Column (1): SFH time. Here, $\tau_x$ denotes the lookback time at which the cumulative stellar mass reaches $x\%$ of the final value at $z=5$. Columns (2) and (3): Mean $\tau_x$ values for the \BG\nspace and \Rad\nspace runs, averaged over the seven target systems. Column (4): Delay in the \Rad\nspace runs relative to \BG.}
\label{tab:sfhtimes}
\end{table}

Table \ref{tab:sfhtimes} summarizes the mean quenching times of the seven target systems in the \BG\nspace and \Rad\nspace runs. Here, $\tau_{x}$ denotes the lookback time at which the cumulative stellar mass reaches $x\%$ of the final stellar mass at $z=5$. We report the values averaged over the seven UFDs rather than individual halo-by-halo values, in order to provide a compact quantity that can be compared with observationally inferred SFH timescales.

The mean values of $\tau_{90}$ and $\tau_{80}$ differ by approximately $440\,\rm{Myr}$ between the two reionization prescriptions. Since the \BG\nspace and \Rad\nspace runs differ only in how reionization is implemented, this comparison isolates the impact of the reionization treatment itself. Thus, the delayed quenching found here suggests that part of the observed diversity in UFD quenching times may arise from variations in the local timing and strength of reionization. Therefore, quenching-time distributions are among the most direct observational diagnostics for testing whether patchy reionization affects UFD evolution.

Of course, reionization is not the only process that determines UFD quenching times. Observed UFDs are affected by a combination of physical processes, including infall into a massive host (e.g., \citealp{Wetzel2015, Fillingham2019, sykim2026}), tidal stripping (e.g., \citealp{Fattahi2018, GarrisonKimmel2019-tidal}), ram-pressure stripping (e.g., \citealp{Simpson_2018, Hausammann2019}), and internal stellar feedback (e.g., \citealp{Brown_2014, Wheeler2015, Jeon_2017}), in addition to the timing and topology of reionization.

Because our simulations are evolved only to $z=5$ and target isolated systems, it is difficult to directly address the impact of later infall into a massive host or the subsequent environmental processing of satellites. Thus, we do not attempt a one-to-one comparison with observed UFDs, many of which may have experienced complex orbital histories. Instead, this study isolates one relevant factor, the treatment of reionization. Even this single change produces a difference of approximately $440\, \rm{Myr}$ in the mean values of $\tau_{80}$ and $\tau_{90}$. Therefore, patchy reionization could be considered an important ingredient when interpreting the observed diversity of UFD quenching times.

\subsection{Implications for previous UFD simulations}

Our \BG--\Rad\nspace comparison provides a useful framework for interpreting previous UFD simulations that adopted homogeneous UVB models. For SFHs, a useful comparison is the work by \citet{Revaz2018}, who modeled dwarf galaxy evolution with a homogeneous \citetalias{HaardtMadau2012} UVB and found that while many observed dwarf properties could be reproduced, gas-rich and still star-forming faint systems such as Leo T and Leo P remained difficult to explain; they explicitly pointed to the need for a better understanding of UVB heating. \citet{Wheeler2015} also adopted \citetalias{FaucherGiguere2009} UVB and found that their UFDs have uniformly ancient stellar populations, which they attributed to reionization quenching. Additionally, in the Marvel/Justice League framework, \citet{Munshi2021} modeled reionization with the \citetalias{HaardtMadau2012} UVB and showed that most UFDs are quenched by reionization. They also noted that the UVB prescription can affect baryon retention and galaxy occupation at the low-mass end. In this context, our \BG--\Rad\nspace comparison suggests that the rapid truncation commonly seen in simulations that adopt a homogeneous UVB is partly caused by the way reionization is implemented. Although our \Rad\nspace runs do not imply that such systems would necessarily avoid quenching, they do show that a more gradual and spatially inhomogeneous radiation field can preserve star-forming gas for longer and thereby delay quenching relative to a spatially uniform UVB treatment.

\par
Regarding the spatial extent of the simulated UFDs, we avoid a definitive interpretation because size measurements at $z=5$ may be affected by the dynamical instability of several systems. Despite this caveat, Figure \ref{fig:sfregion} suggests a potential correlation between reionization and galaxy size; in most halos, in-situ star formation following the onset of reionization is more centrally concentrated than the pre-reionization component, typically occurring within $\sim 0.3\,R_{\rm{vir}}$. This indicates that cosmic reionization may influence the spatial distribution of stellar populations in low-mass systems, although our current measurements are insufficient to conclude that RT-based reionization directly leads to increased compactness.

\par
\subsection{Caveats}
\label{caveat}
We should note that the results presented here are subject to several numerical and physical caveats, which we outline below.

\begin{itemize}
    \item \textbf{Scope and limitations of the reionization setup}
    
    \begin{figure}
        \centering
        \includegraphics[width=1.0\linewidth]{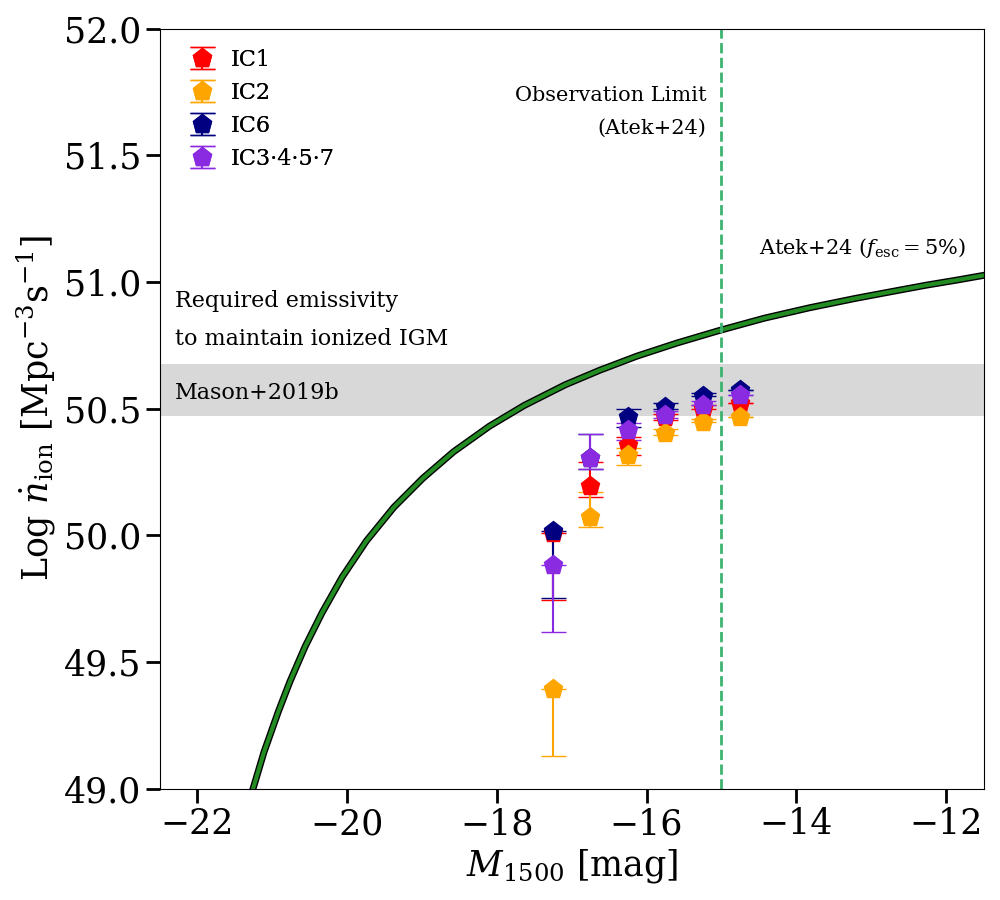}
        \caption{Ionizing emissivity of galaxies at $z\sim7$. Pentagon symbols show the ionizing emissivity inferred from our 40 galaxy sample; error bars indicate the 16th-84th percentiles. Colors denote different initial conditions. The gray band marks the emissivity required to maintain an ionized IGM at $\simeq 7$, following \citet{Mason2019b}. The green line corresponds to the emissivity inferred using $f_{\rm esc}=5\%$, reported in \citet{Atek2024}. The green vertical dashed line indicates the observational limit adopted by \citet{Atek2024}. Our inferred emissivities lie near the lower edge of the required-emissivity band from \citet{Mason2019b} and are $\sim 0.25$ dex below the emissivity shown in \cite{Atek2024}.
        }
        \label{fig:ionizingemissivity}
    \end{figure}

    We adopt a parent volume of $L_{\rm box} = 6.25\,h^{-1}\,{\rm cMpc}$ to maintain the resolution of the DM mass required to resolve the UFD progenitors $z=0$ ($M_{\rm vir} \sim 10^{8} \Msun$). Consequently, we employ this modest box size as a necessary trade-off to simultaneously select zoom-in targets and construct the reionization source population. Furthermore, we limit the number of ionizing sources to the 40 most massive halos to manage the computational expense of the RT calculation, as additional sources would increase the number of photon packets.

    \par
    These choices imply that our source population is designed to investigate local radiative feedback rather than to implement global reionization. Our setup should therefore be interpreted as a controlled model for the local radiative influence of nearby systems, rather than a representation of the global ionizing emissivity. Figure \ref{fig:ionizingemissivity} compares the ionizing emissivity of our 40 sources with observational and theoretical predictions. The pentagon symbols represent our source population (with a range of 16 to 84\%), while the gray band denotes the emissivity required to maintain an ionized IGM at $z \simeq 7$ (\citealp{Mason2019b}). Although our emissivities are generally lower than the values inferred by \citet{Atek2024} (green line), including galaxies down to $M_{\rm{1500}} \leq -15$ brings our emissivity to a level comparable to that required to maintain reionization. We emphasize that this comparison pertains to maintaining an ionized state, not driving reionization from a neutral IGM. Nevertheless, these results confirm that our setup is sufficient to capture local reionization effects within the zoom-in regions.

    \item \textbf{Choice of the activation redshift} \\
    Both schemes activate reionization at $z=7$, which presumes that the zoom-in volumes are still largely neutral at that epoch even though the global ionized fraction is already  $Q_{\rm HII} \approx 0.5$ by then (e.g., \citealp{Planck2018, Mason2018, Davies2018}). We verified that this is the case. Using the excursion set description of reionization (\citealp{Furlanetto2004}), we computed the local reionization redshift, $z_{\rm reion}$, at the location of each zoom-in region within its parent DM-only volume. The calculation is calibrated to reproduce the observed global reionization history. We find $z_{\rm reion}=6.59-7.06$ for the seven regions, meaning that each reionizes at or after the midpoint of its parent box. Such late reionization is the expected behavior for the isolated, underdense environments in which our targets reside, and it agrees with the environment-dependent reionization epochs measured in the full RHD simulation THESAN \citep{Zhao2026} for comparable overdensities. Moreover, the \Rad\nspace runs produce this timing self-consistently, reaching half ionization at $z=6.10-7.11$ (Figure \ref{fig:volumefillingfraction}). The activation at $z=7$ therefore reflects the physical reionization epoch of these environments rather than a numerical convenience.

    \item \textbf{Restriction to the reionization era} \\
    We emphasize that our simulations are evolved only to $z=5$, the epoch by which hydrogen reionization is observationally inferred to be largely complete (e.g., \citealp{Becker2015, Mcgreer2015, Bosman2022}). Our setup therefore does not capture late-time mergers or any rejuvenation of star formation after $z=5$. Below this redshift, the meta-galactic ionizing background is increasingly governed by distant massive star-forming galaxies and AGNs (e.g., \citealp{Faucher_Gigu_re_2020, Finkelstein2022}), neither of which is represented in our source model.
    
    A direct consequence of this scope is that five \Rad\nspace halos (HALO1, HALO2, HALO3, HALO5, and HALO6) are still forming stars at $z=5$, so their $\tau_{90}$ values, and the $\sim 440\,\rm{Myr}$ mean delay, should therefore be regarded as lower limits for our current source model. Capturing the remainder of their star formation would require following them past $z=5$.  Simply extending the present setup, however, would not be physical: with only nearby star-forming galaxies as sources, our radiation field misses the distant galaxies and AGNs that dominate the ionizing background at lower redshifts, and the runs would artificially prolong star formation. A realistic continuation would instead replace our local sources with a uniform cosmic UVB that represents these distant contributions.

    Such a continuation has already been carried out by \citet{Kim2023} within the same framework. They modeled reionization with an environment-dependent UVB computed from nearby galaxies and then switched to the uniform UVB (\citetalias{HaardtMadau2012}) at $z=5.8$, once distant sources are expected to dominate. They found that star formation in their UFD analogs ($M_{\rm vir} \sim 7.6\times10^{7}-1.6\times 10^{8}\Msun$ at $z=7$) ceases almost immediately once the background is imposed. Given that the halos still forming stars in our runs are lighter than these systems, even at the later epoch of $z=5$ ($M_{\rm vir} \sim 2-5\times10^{7}\Msun$), they would quench at least as quickly under such a background. As an independent rough check, we computed a star-forming gas depletion time at $z=5$, defined as $t_{\rm{SF\, dep}}=M_{\rm{SF\,gas}} / \rm{SFR}$. This simple estimate gives exhaustion by $z \sim 4$ for four of the five \Rad\nspace halos; HALO6 is excluded because its low SFR at $z=5\,(\sim 10^{-6} \Msun\,\rm{yr}^{-1})$ renders its depletion time uninformative as a remaining star forming lifetime. Taken together, the ${440}\,\rm{Myr}$ delay is formally a lower limit within our source model, but the behavior found by \citet{Kim2023} implies that extending the sources to lower redshift would not increase it much and the true delay is likely close to $\sim 440\,\rm{Myr}$.

    \item \textbf{Simplified escape fraction model} \\
    We adopt a single redshift-dependent escape fraction (\citealp{Faucher_Gigu_re_2020}, $f_{\rm esc} \simeq 0.03-0.05$) applied identically to all sources and to both schemes. The escape fraction of a real galaxy, however, depends on its halo mass, environment, and the burstiness of its star formation. Its ionizing radiation also likely leaks anisotropically, through a small number of cleared channels. Our framework cannot capture these dependences: the sources exist only in the DM-only parent run and are represented as point sources, so they carry no information about their ISM or escape geometry. Simulations that measure $f_{\rm esc}$ of galaxies do not converge on these dependences either. At our source masses, the reported values span roughly 5 to 20 percent (e.g., \citealp{Xu2016, Kimm&Cen2014, Ma2020fesc, Rosdahl2022}), which places our adopted  \citetalias{Faucher_Gigu_re_2020} values at the low end of this range. To test the sensitivity of our main result to this choice, we reran HALO2-\Rad\nspace with $f_{\rm esc}=20$ percent, the highest value reported at our source masses. We confirmed that the halo still sustains star formation down to $z=5$, and its quenching delay relative to the BG counterpart changes only from 480 to about 430 Myr. The delayed quenching found in this work is therefore not sensitive to the adopted low escape fraction.

    \item \textbf{Omission of internal Pop~II radiative feedback} \\
    Our simulations include on-the-fly RT only for Pop~III stars formed within the zoom-in region and the corresponding Pop II radiation is not transported, for computational efficiency. Although Pop~II stars dominate the cosmic ionizing budget during the EoR, the internal budget of our zoom-in regions need not follow the global one. Each volume is small and selected to contain an isolated UFD with no massive halos nearby. In such underdense regions, star formation is weak and metal enrichment proceeds slowly, so pristine gas survives and Pop~III stars continue to form in the surrounding minihalos even during the EoR. We quantified the omitted Pop~II RT channel by estimating the cumulative ionizing energy of each stellar population. Over the full simulated history, the omitted internal Pop~II energy amounts to $\sim$ 5 percent of the Pop~III energy that we do include, and $\sim$ 4 percent when the integration is restricted to $5 \leq z \leq 7$. Compared with the external reionization field over the same interval, the internal Pop~II RT channel is smaller by factors of $\sim 3-373$ , and the external field remains dominant in all seven regions. The omitted channel is thus subdominant both to the included Pop III RT and to the external UV field.

    \item \textbf{Limited statistical sample} \\
    Our sample consists of seven halos, and halo-to-halo scatter is known to be large in this mass regime. We therefore do not treat the mean delay of $\sim 440\,\rm{Myr}$ as a converged statistic. Nor do we treat the fractions of halos with delayed quenching (six of seven) or with star formation sustained to $z=5$ (five of seven) as population fractions. The individual delays themselves range from $0$ to $740\,\rm{Myr}$ (Table \ref{tab:tau90}). What our sample can suggest is the qualitative conclusion that the reionization treatment changes when a UFD quenches. Each halo is simulated with both schemes from identical initial conditions, so a delay measured within a pair can only come from the reionization treatment. We accordingly suggest the $\sim 440\,\rm{Myr}$ mean as the characteristic size of the effect, comparable to the delays inferred for observed UFDs (\citealp{Sacchi_2021, Meredith2025}).

    A natural question is whether, despite the small sample size, our seven halos are representative of the UFD population rather than unusual systems. In the MZR (Figure \ref{fig:MZR}), our halos fall within the range of previous UFD simulations, and the known metallicity scatter of this regime is present within our sample. We also tested their positions in the SMHM plane. Extrapolating to $z=0$ halo masses with the stellar masses fixed at their $z=5$ values, our systems fall within the $1-2\sigma$ ranges of  relations proposed by previous studies (\citealp{Behroozi2019, Nadler2020}) and overlap the region occupied by previous UFD simulations. We find no indication that our sample is biased toward unusual systems. This check addresses selection bias, however, not statistical convergence: establishing converged delay statistics requires a larger sample, which we leave to future work.

\end{itemize}

\section{Summary and Conclusions}
\label{Sec:Summary}
This work investigates the impact of reionization modeling on the evolution of seven ultra-faint dwarf galaxies (UFDs) using high-resolution cosmological zoom-in simulations ($m_{\rm{gas}} \approx 63\,\Msun$). We compare two distinct implementations derived from identical ionizing sources and spectra: (i) a spatially uniform, redshift-dependent UV background (\BG) and (ii) an on-the-fly radiative transfer (\Rad) model that tracks the time-dependent propagation of ionizing photons. By isolating the effects of reionization modeling, we evaluate its influence on the thermodynamic properties of the IGM, the evolution of star-forming gas, and the resulting star formation histories (SFHs). Finally, we identify potential observable signatures that may distinguish between these two reionization prescriptions.

\vspace{\baselineskip}

Our main findings are summarized as follows.

\paragraph{The zoom-in gas properties}
The gas properties show the clearest difference between the two reionization models. In \BG, the gas is heated almost immediately once reionization begins at $z=7$, and the mean gas temperature in the zoom-in region reaches $\langle T_{\rm gas,\,BG} \rangle \approx 10^{3.7}\,\rm K$. In \Rad, by contrast, ionization fronts sweep across the region gradually, so the gas remains cooler at the same epoch, with $\langle T_{\rm gas,\,Rad} \rangle \approx 10^{2.9}\,\rm K$ at $z=7$. The dense clumps with $n_{\rm H}>10\,\rm cm^{-3}$ disappear by $z\approx 6.0$ in \BG, whereas \Rad\nspace retains such clumps until $z=5$. The ionized volume filling fraction shows the same behavior: \BG\nspace jumps to $Q_{\rm HII}=1$ at $z=7$, while \Rad\nspace increases more gradually over time.

\paragraph{Star formation history}
The different gas evolutions lead to markedly different SFHs. In \BG, the cumulative SFHs increase until around $z=7$ and then flatten in all seven halos. Six of the seven halos show delayed quenching under \Rad\nspace ($\Delta\tau_{90}=300-740\,\rm{Myr}$), and five of them continue forming stars down to $z=5$. The extension in HALO4 is limited to a brief episode at $z \approx 7.0$, and HALO7, quenched before the onset of reionization, shows no delay. Using the timescale at which 90\% of the final stellar mass has formed, we find that quenching in \Rad\nspace is delayed by 440\,Myr on average relative to \BG. 

\paragraph{Star-forming region}
Reionization modeling also changes where star formation occurs within the halo. After reionization begins, star formation in the \Rad\nspace runs becomes more centrally concentrated. At $z<7$, about 99\% of all in-situ star formation events in \Rad\nspace occur within $0.3\,R_{\rm vir}$, while only $\sim$1\% occur outside this radius across the seven halos. This trend arises because the \Rad\nspace prescription effectively shields and preserves dense gas in the halo centers, whereas the outer regions are heated up by ``outside-in" ionizing radiation. Conversely, the \BG\nspace model rapidly and uniformly evacuates star-forming gas from the entire zoom-in region soon after the UVB is switched on.

\paragraph{Internal quenching}
We find that the response to the \Rad\nspace prescription is not uniform across all systems. While five halos exhibit extended SFHs down to $z=5$, HALO4 and HALO7 fail to do so: neither system re-enters a sustained star-forming phase after early internal feedback. In HALO4-\Rad, an enhanced period of Pop III star formation at $z\sim10$—about four times the rate of other systems—drives significant gas evacuation. Although star formation briefly reignites after $\sim 300\,\rm Myr$, the resulting core-collapse supernova (CCSN) feedback removes the remaining dense gas, preventing further recovery. For HALO7-\Rad, the most massive system in our sample, the combined impact of frequent Pop II CCSNe and the onset of reionization at $z=7$ leads to complete quenching. These cases demonstrate that the evolutionary trajectory of a UFD can be decisively shaped by the stochasticity of internal feedback.

\paragraph{Stellar metallicity}
Because our halos follow various evolutionary pathways, the stellar mass--metallicity relationship does not cleanly separate \BG\nspace and \Rad\nspace into two distinct sequences. Instead, both sets of runs scatter across the plane. Notably, the four \Rad\nspace halos that exceed $10^{4}\,\Msun$---HALO2, HALO3, HALO5, and HALO7--- occupy a consistent metallicity range of $-3.0 \lesssim \langle [\rm Fe/H] \rangle \lesssim -2.5$, aligning with existing theoretical and observational studies. However, in line with other existing models, our simulations also fail to reproduce the observed population characterized by lower stellar masses ($M_{\star} \lesssim 10^{4}\,\Msun$) and higher metallicities ($\langle [\rm Fe/H] \rangle \gtrsim -2.5$).

\paragraph{Carbon and magnesium abundance}
Carbon provides the clearest abundance signature of extended star formation. In the \Rad\nspace runs, HALO2 and HALO5 show post-reionization [C/Fe] values about 0.5 dex higher than those of their pre-reionization stellar populations, likely driven by Pop II AGB enrichment during extended star formation. We do not find a clear knee in the [Mg/Fe]-[Fe/H] plane. However, HALO5-\Rad\nspace contains stars with elevated [Mg/Fe], which can plausibly be explained by enrichment from a top-heavier Pop II CCSN progenitor population. In general, the abundance patterns of our simulated halos are broadly consistent with those observed in Local Group UFDs.

\paragraph{Detectability as DLAs}
We further investigate whether \Rad \nspace halos retain a neutral gas sufficient to be detectable as damped Lyman-$\alpha$ absorbers (DLAs), defined by $N_{\rm HI} \ge 10^{20.3}\,\rm cm^{-2}$ \citep{Wolfe1986}. In \BG, DLA-level gas survives to $z=5$ in only three of the seven halos, with areas below $0.1 \, \rm{kpc}^{2}$ whereas in \Rad\nspace it survives in six halos with areas up to $\sim 0.4\, \rm{kpc}^{2}$. These cross-sections are too small for practical detection, but the systematically larger area in \Rad\nspace indicates that RT-based reionization preserves more neutral gas at the DLA level in low-mass systems.

\vspace{\baselineskip}

Overall, our results demonstrate that the numerical implementation of reionization qualitatively changes the evolution of UFDs, even when ionizing sources and spectra are held constant. Compared with a spatially uniform UVB, RT-based patchy reionization helps to survive the dense gas, thus delaying or weakening quenching in most of our simulated systems. These differences leave measurable signatures in their SFHs and neutral gas content, supporting the necessity of realistic reionization modeling to explain the observed diversity of Local Group UFDs.

However, while RT-based reionization tends to promote more extended star formation, our findings suggest that caution is required when interpreting other observables as definitive evidence of patchy reionization. As shown in our analysis, even when extended star formation is present, it does not necessarily leave a distinct, identifiable signature in the stellar metallicity distribution. This lack of a clear chemical imprint likely arises because the delay in quenching—while physically significant—is not sufficiently prolonged to allow for substantial chemical enrichment. Nevertheless, upcoming wide and deep surveys, such as the Legacy Survey of Space and Time (LSST) and the Nancy Grace Roman Space Telescope, will significantly expand the census of Local Group and field UFDs. These future observations will provide increasingly powerful tests of whether patchy reionization has left observable imprints on the evolutionary diversity of the faintest galaxies.

\section*{acknowledgements}
The authors also thank Volker Springel, Joop Schaye, and Claudio Dalla Vecchia for granting permission to use their versions of \textsc{gadget}. M.~J. acknowledges support from the Korean government-funded National Research Foundation (NRF) grant (MSIT) under the grant number 2022M3K3A1093827. M.~J. was also supported by Samsung Science and Technology Foundation under Project Number SSTF-BA2402-03. The work of Y. Choi is supported by NSF NOIRLab, which is managed by AURA under a cooperative agreement with the U.S. National Science Foundation. Part of this work was performed on the Nurion supercomputer at the Korea Institute of Science and Technology Information (KISTI), supported by the National Supercomputing Center (KSC-2025-CRE-0078).


\bibliography{myrefs}
\bibliographystyle{aasjournal}

\end{document}